\documentclass[pdflatex,sn-mathphys-num,iicol]{sn-jnl}

\usepackage{graphicx}%
\usepackage{multirow}%
\usepackage{amsmath,amssymb,amsfonts}%
\usepackage{amsthm}%
\usepackage{mathrsfs}%
\usepackage[title]{appendix}%
\usepackage{xcolor}%
\usepackage{textcomp}%
\usepackage{manyfoot}%
\usepackage{booktabs}%
\usepackage{algorithm}%
\usepackage{algorithmicx}%
\usepackage{algpseudocode}%
\usepackage{listings}%
\usepackage{float}
\usepackage{amssymb} 
\usepackage{subfigure}
\usepackage{xcolor}
\usepackage{graphicx}
\usepackage{dcolumn}
\usepackage{bm}
\usepackage{physics}
\usepackage{siunitx}
\usepackage{ulem}
\usepackage{mathrsfs}
\usepackage[mathlines]{lineno}  
\renewcommand{\vec}[1]{\mathbf{#1}}

\newcounter{saveenum}
\begin{document}

\title[Entanglement-enabled image transmission through complex media]{Entanglement-enabled image transmission through complex media}

\author*[1]{\fnm{Chloé} \sur{Vernière}}\email{chloe.verniere@insp.jussieu.fr}

\author[1]{\fnm{Raphaël} \sur{Guitter}}

\author[1,2]{\fnm{Baptiste} \sur{Courme}}

\author*[1]{\fnm{Hugo} \sur{Defienne}}\email{hugo.defienne@insp.jussieu.fr}

\affil[1]{Sorbonne Université, CNRS, Institut des NanoSciences de Paris, INSP, F-75005 Paris, France}

\affil[2]{\orgdiv{Laboratoire Kastler Brossel}, \orgname{ENS-Universite PSL, CNRS, Sorbonne Universite, College de France}, \orgaddress{24 rue Lhomond}, 75005 Paris, France }

\abstract{Scattering in complex media scrambles light, thus obscuring images and limiting applications from astronomy to microscopy. Existing computational and wavefront-shaping methods treat scattering as a linear optical-wave inversion problem that aims to render the medium transparent by inverting the scattering process. As classical approaches, they do not account for the quantum nature of the incident field. Here, we demonstrate a quantum-entanglement-based method that enables selective image transmission through complex media. The medium is effectively turned into a quantum-classical image filter via wavefront shaping — images encoded on an entangled two-photon state are transmitted faithfully, while those carried by classical light remain fully scattered and unreadable. This method exploits a property of quantum entanglement — the preservation of photon correlations across multiple measurement bases — that has no classical counterpart. Therefore, we establish an approach for controlling light in complex media by tailoring solutions to the quantum properties of the input state, with potential applications in secure information transmission by rendering channels opaque to classical signals while preserving the quantum link.}

\maketitle

Transmitting optical spatial information, such as the image of an object, with high fidelity is crucial for numerous applications in optics. Yet, it becomes highly challenging when light propagates through complex disordered media, such as biological tissues, turbulent atmosphere, or multimode fibers, where spatial information is scrambled and the image becomes unreadable~\cite{davies_adaptive_2012,ntziachristos_going_2010}. In recent years, efforts to overcome scattering and enable imaging through complex media have led to the development of many techniques~\cite{Bertolotti2022}.

Among these, wavefront shaping has emerged as a powerful tool~\cite{Cao2022,rotter_light_2017}. In this context, spatial light modulators (SLMs) can be used, for example, to focus light through a complex medium by shaping its wavefront at the input~\cite{vellekoop_focusing_2007}.
In non-invasive scenarios, where access to the far side or inside of the medium is not possible, the wavefront correction applied via the SLM can be retrieved for example by optimizing guidestars~\cite{vellekoop_focusing_2007,horstmeyer_guidestar-assisted_2015,katz_looking_2012}, image quality metrics~\cite{yeminy_guidestar-free_2021,haim_image-guided_2025,daniel_light_2019,boniface_non-invasive_2020}, or via medium modeling~\cite{thendiyammal_model-based_2020}.
Once the correction is known, wide-field imaging becomes possible within a limited field of view (FOV), allowing light to be steered through the medium to form an image as if it were transparent.
This FOV is limited by the medium’s memory effect~\cite{freund_memory_1988}, but can be extended via multiple measurements~\cite{badon_distortion_2020,boniface_noninvasive_2019} and multi-plane wavefront correction~\cite{kupianskyi_all-optically_2024}.

Alternatively, computational imaging methods can bypass physical correction performed by an SLM. If part of the scattering matrix of the medium is known~\cite{gigan_matrice_2010}, it is then possible to reconstruct an object hidden behind the medium from a measured output field or intensity using linear inverse problem algorithms~\cite{Popoff2010,liutkus_imaging_2014}. Even without knowing the matrix, its intrinsic correlations can be exploited to reconstruct an object. For example, provided the object lies within the memory effect range of the medium, direct reconstruction from speckle patterns is achievable using phase retrieval algorithms~\cite{Bertolotti2012,Katz2014} and bispectrum analysis~\cite{wu_single-shot_2016}. Under similar experimental conditions, machine learning approaches have also been used to retrieve hidden objects from speckle image datasets~\cite{li_imaging_2018,li_deep_2018}.

Despite their diversity, all these techniques share a common principle: restoring imaging by inverting the scattering process. Computational approaches estimate the inverse of the scattering matrix to reconstruct the input field, while wavefront shaping methods physically implement this inverse. 
Both are based on the same fundamental physics - linear wave propagation in complex media - which is valid in nearly all practical situations and described by a unified framework~\cite{gigan_matrice_2010}:
\begin{equation}
\label{eq1}
E^{\text{out}} = S E^{\text{in}},
\end{equation}
where $E^{\text{in}}$ and $E^{\text{out}}$ are vectors representing the input and output fields, and $S$ is the scattering matrix encompassing both the complex medium and the imaging optics.

\begin{figure} [h!]
    \centering
    \includegraphics[width=0.99\linewidth]{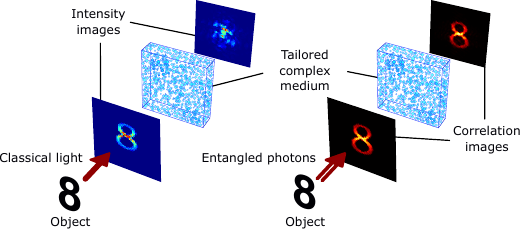}
    \caption{\textbf{Rendering a complex medium transparent to {entangled photons} while remaining opaque to classical light.} Two imaging scenarios through a complex medium are shown. On the left, an object encoded in classical light intensity is scattered, producing a speckle pattern at the output. On the right, the object is encoded in the spatial correlations of entangled photon pairs. Despite propagation through the same medium, the measured correlation image reveals the object as if the medium were transparent. This is achieved by tailoring the optical disorder of the medium for implementing a specific basis transformation that preserves quantum correlations and thus the encoded image. Both intensity and correlation images are from numerical simulations detailed in Methods and Supplementary Section IV.}
    \label{fig:Fig1}
\end{figure}

\begin{figure*} [hbt!]
    \centering
    \includegraphics[width=0.99\textwidth]{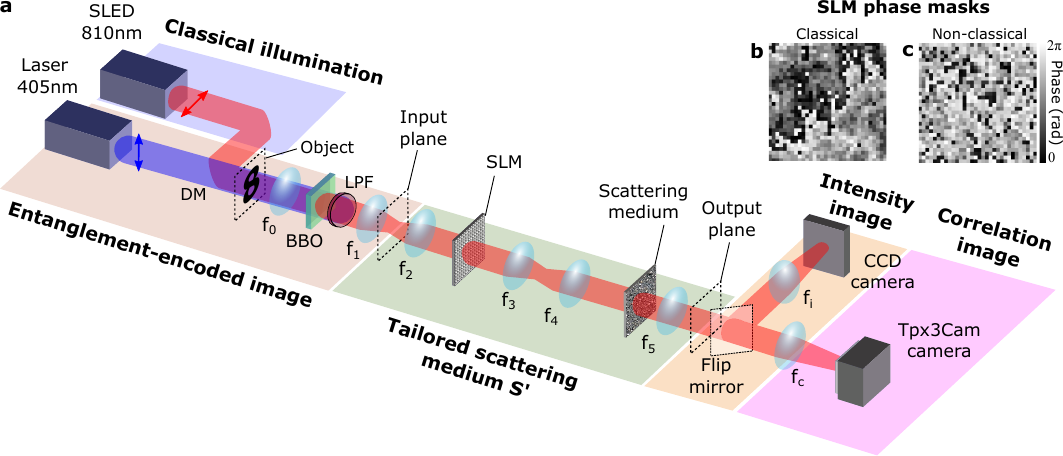}
    \caption{\textbf{Experimental setup.} \textbf{a,} A continuous-wave laser at 405nm illuminates an object placed in the object plane of a lens $f_0$. The beam is then focused onto a 0.5mm-thick nonlinear $\beta$-barium borate (BBO) crystal, where type-I spontaneous parametric down conversion (SPDC) generates degenerate spatially entangled photon pairs at 810nm. A long-pass filter (LPF) removes the residual pump beam after the crystal. The photon pairs are Fourier-imaged by another lens $f_1$ onto the input plane, which is conjugate to the object plane. Before the object, a dichroic mirror (DM) combines the paths of a superluminescent diode (SLED) and the pump laser. The lasers are switched on or off depending on whether the system operates under quantum or classical illumination.
    Two confocal telescopes formed by the lenses $f_2$–$f_5$ link the input plane to the output plane. A spatial light modulator (SLM) and a scattering medium (a layer of Parafilm) are positioned in the Fourier planes of the telescopes. The optical transformation $S'$ linking the input plane to the output plane is tailored by programming phase patterns onto the SLM  {using either a \textbf{(b)} classical or \textbf{(c)} our non-classical wavefront-shaping approach (see Fig.~\ref{fig:Fig3} for details).}
    Lenses {$f_i$ and $f_c$}, along with movable mirrors, {represent} two separate imaging systems {with magnification $1$}, mapping the output plane onto a Charge-Coupled Device (CCD) camera and a time-stamping Tpx3Cam, respectively. A bandpass filter at $810\pm5$nm (not shown) is used to suppress stray light on both cameras.}
    \label{fig:Fig2a}
\end{figure*}

\begin{figure*} [hbt!]
    \centering
    \includegraphics[width=0.99\textwidth]{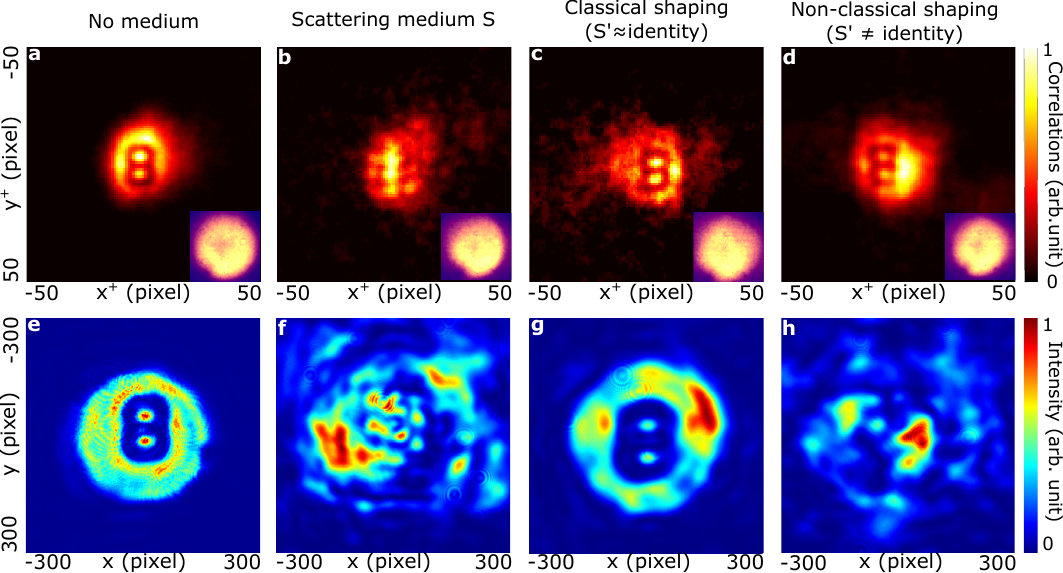}
    \caption{\textbf{Experimental results.} In the quantum case, correlation images of an entanglement-encoded object (the digit `8') are measured \textbf{(a)} without the scattering medium, \textbf{(b)} after propagation through it, after tailoring the medium via  {\textbf{(c)} classical and \textbf{(d)} non-classical wavefront shaping}. The corresponding intensity images, shown in the insets, reveal no information about the object.
    In the classical case, intensity images are recorded \textbf{(e)} without the scattering medium, \textbf{(f)} after propagation through it, after tailoring it via  {\textbf{(g)} classical and \textbf{(h)} non-classical wavefront shaping}. The same scattering medium and tailoring function (i.e. the same SLM phase pattern) are used in both quantum and classical configurations. Correlations images required $8$-second acquisitions, while classical intensity images were acquired withing a few tens of milliseconds. Each image is normalized to its maximum value. See Supplementary Section I for more details about the experimental setup.}
    \label{fig:Fig2b}
\end{figure*}

Paradoxically, the simplicity of Equation~\eqref{eq1} is also its main limitation, as it does not account for the quantum nature of the field. The dual linearity of quantum optics - of Maxwell’s equations and quantum mechanics - requires a broader framework to fully describe the propagation and properties of non-classical states~\cite{fabre_modes_2020}. 
For example, when entangled photon pairs propagate through disordered media, they undergo complex interference processes producing  {unique} output correlation patterns known as two-photon speckles~\cite{beenakker_two-photon_2009,peeters_observation_2010,defienne_two-photon_2016,defienne_adaptive_2018,CourmeCertifcation,lib_real-time_2020,courme2025nonclassicaloptimizationcomplexmedia,safadi_coherent_2023,lib_quantum_2022}. Unlike the classical case, the output two-photon field depends bilinearly on the scattering matrix:
\begin{equation}
\label{eq2}
\Psi^{\text{out}} = S \Psi^{\text{in}} S^t,
\end{equation}
where $S^t$ is the transpose of the scattering matrix, and $\Psi^{\text{in}}$ and $\Psi^{\text{out}}$ are matrices representing the discrete input and output two-photon wavefunctions, respectively (see Methods). This quantum framework is fundamentally richer than its classical counterpart, offering access to solutions that are inaccessible using classical light. 

In this work, we introduce an approach for image transmission through complex media that harnesses the physics of non-classical wave propagation.
As illustrated in Figure~\ref{fig:Fig1}, we demonstrate that the medium's disorder can be tailored to become transparent to entangled photon pairs, while remaining opaque under classical illumination.
We exploit this concept to experimentally transmit images of arbitrary objects encoded in the photons spatial correlations, under conditions where classical intensity-based imaging fails.

\section*{Experimental results}

Figure~\ref{fig:Fig2a}a shows the experimental setup. In the quantum case, images of arbitrary objects are encoded in the spatial correlations of entangled photon pairs using the method described in Ref.~\cite{HidingImages}. For that, the object is illuminated with a 405 nm blue pump beam, which is Fourier-transformed by a lens onto a type-I Beta Barium Borate (BBO) crystal. This crystal generates frequency-degenerate (810 nm) spatially-entangled photon pairs via spontaneous parametric down-conversion (SPDC).
In the classical case, the object is illuminated with a superluminescent diode (SLED) centered at 810 nm. The light then propagates through a tailored complex medium, described by a matrix $S'$, consisting of a SLM and a scattering medium (a parafilm layer), both placed in planes close to the Fourier plane of the object.
At the output, a Charge Coupled Device (CCD) camera measures intensity images in the classical case, while a Tpx3Cam camera records correlation images in the quantum case.

{When no medium is present, Figure~\ref{fig:Fig2b}a shows the correlation image of a binary digit `8' object obtained using the quantum-encoding scheme. As detailed in Methods, this image, denoted $\Gamma_+$, is obtained by measuring the second-order correlation function with the Tpx3Cam camera~\cite{defienne_general_2018,courme_quantifying_2023} and projecting it along the sum-coordinate axis $\vec{r_i}+\vec{r_s}$.
After propagation through the scattering medium, the spatial information becomes scrambled, resulting in a fully diffused object in the output correlation image (Fig.\ref{fig:Fig2b}b). Illuminating the same object with a classical source (Fig.\ref{fig:Fig2b}e)  also produces a speckle intensity pattern at the output (Fig.\ref{fig:Fig2b}f). 
{Under these conditions, classical wavefront shaping can be used to program the SLM (Fig.~\ref{fig:Fig2a}b) and recover the image of the object at the output under both quantum (Fig.~\ref{fig:Fig2b}c) and classical illumination (Fig.~\ref{fig:Fig2b}g). In this case, the overall transformation $S'$, corresponding to the combination of the scattering medium and the SLM phase pattern, is close to the identity (see Supplementary Section~XV), thereby effectively rendering the medium transparent.  {We refer to the reconfiguration of the initial matrix $S$ (flat SLM + medium) into the target matrix $S'$ (programmed SLM +medium) as `tailoring the disorder'}.

Interestingly, beyond classical wavefront shaping solutions, there is a distinct class of solutions - referred to here as non-classical solutions - that is transparent only to quantum correlations. For example, programming the SLM with the phase pattern shown in Fig.~\ref{fig:Fig2a}c allows images encoded in the spatial correlations of entangled photon pairs to be transmitted (Fig.~\ref{fig:Fig2b}d), while classically encoded images are not transmitted and remain completely blurred (Fig.~\ref{fig:Fig2b}h). This solution is not predicted by Eq.~\eqref{eq1}: it} implements a transformation $S'$ that differs from both the original $S$ and the identity, yet nonetheless preserves the photons’ spatial quantum correlations and therefore the encoded image. 

\section*{Tailoring optical disorder}

\begin{figure}
    \centering
    \includegraphics[width=0.99\linewidth]{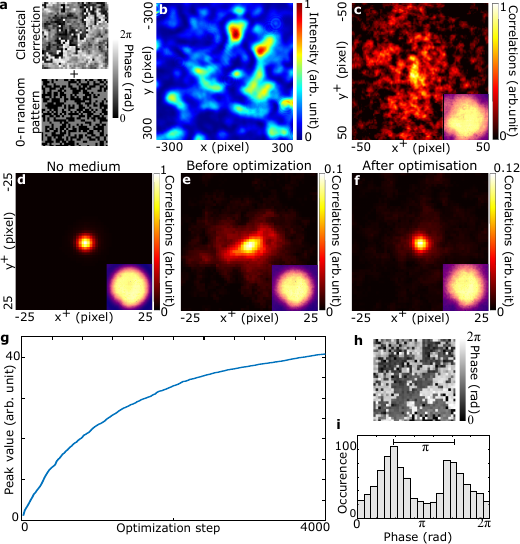}
    \caption{\textbf{Optimization process.} {The sum of the classical correction mask (Fig.~\ref{fig:Fig2a}b) and a random $0$/$\pi$ mask \textbf{(a)} constitutes an exact solution of Eq.~\eqref{eq3ter}, while preserving neither the classically encoded image \textbf{(b)} nor the entanglement-encoded image \textbf{(c)}. \textbf{d-f},} Correlation images measured experimentally using a maximally entangled two-photon state at the input (guide state) without the scattering medium \textbf{(d)}, with the scattering medium and before optimization \textbf{(e)}, and after optimization \textbf{(f)}. \textbf{g,} Optimization curve, showing the variation of the central value in the correlation image in function of the number of steps. The optimization is performed computationally by simulating the experiment on a computer using experimentally measured transmission matrix and two-photon input state. The optimization converges towards an SLM phase mask{, shown in Figure~\ref{fig:Fig2a}c}, that is then used experimentally to tailor the scattering medium. {\textbf{h}, Phase difference between the non-classical (Fig.~\ref{fig:Fig2a}c) and classical masks (Fig.~\ref{fig:Fig2a}b). \textbf{i}, Statistical distribution of the phase values in the difference pattern.}}
    \label{fig:Fig3}
\end{figure}

To understand how {such a non-trivial transformation} $S'$ is implemented, we first examine the two-photon wavefunction {in the input plane}:
\begin{equation}
\label{eq3}
{\psi^\text{in}}({\vec{r_i},\vec{r_s}}) = t\left(\frac{\vec{r_i}+\vec{r_s}}{M} \right) \exp{\frac{-|\vec{r_i}-\vec{r_s}|^2 \pi^2 \sigma_{\vec{r}}^2 }{4 \lambda_p^2 f_1^2}},
\end{equation}
where $\vec{r_i}$ and $\vec{r_s}$ are the idler and signal photons transverse positions, $M = 2 f_1/f_0$ is a magnification factor, ${t}$ is the encoded object function {evaluated at $(\vec{r}_i + \vec{r}_s)/M$}, $\sigma_{\vec{r}} = \sqrt{ L \lambda_p/(12 \pi)}$ is the position correlation width, $\lambda_p$ the pump wavelength in the crystal and $L$ its thickness 
{(derivation of Equation~\eqref{eq3} is provided in Supplementary Section III, based on the general formalism of Ref.~\cite{Schneeloch_2016})}. {Then, we make the assumption that $L$ is sufficiently small such that $\sigma_{\vec{r}} \approx 0$. The output wavefunction after propagation through the optical transformation $S'$ simplifies as:}
\begin{equation}
\label{eq3bis}
    \psi^{\text{out}}({\vec{r_i'}, \vec{r_s'}}) = \int t(\vec{r_+}) H_{s'}(\vec{r_i'}, \vec{r_s'}, {\vec{r_+}}) d \vec{r_+} ,
\end{equation}
where $H_{s'}(\vec{r_i'}, \vec{r_s'}, {\vec{r_+}}) = \int s'(\vec{r_i'}, M \vec{r_+}/2+\vec{r_-}) s'(\vec{r_s'}, M \vec{r_+}/2-\vec{r_-}) d\vec{r_-}$, $s'$ is the coherent point spread function (PSF) associated with $S'$, $\vec{r_i'}$ and $\vec{r_s'}$ are idler and signal photons transverse positions in the output plane, respectively. 
Under this assumption, a class of non-trivial functions $s'$ can be found that restore the object image at the output, and are written as:
\begin{equation}
    \label{eq3ter}
   s_g'(\vec{r'}, \vec{r}) = \mathcal{F} \{ \operatorname{sign}(g) \}(\vec{r'} - \vec{r}), 
\end{equation}
where $\mathcal{F}$ is the Fourier transform, $g(\vec{r}) \in \mathbb{R}$ is an arbitrary function and $\operatorname{sign}(g) = 1$ if $g>0$, and $-1$ otherwise.
Each solution is thus a function of $\vec{r} - \vec{r}'$ whose Fourier transform is a discontinuous function taking values 1 and -1 (i.e. phases 0 and $\pi$) randomly distributed in the transverse plane. By selecting a function $g$ and applying Equation~\eqref{eq3ter}, one can construct a non-trivial optical transformation $S'$ that renders the system transparent to quantum-encoded images. Because $g$ can be chosen arbitrarily, this approach yields an infinite number of possible solutions. In principle, implementing an arbitrary transformation $S \rightarrow S'$ generally requires multi-plane wavefront shaping~\cite{kupianskyi_all-optically_2024}, although in the thin-scattering regime relevant to our experiment, a single SLM already provides substantial control  (see Methods and Supplementary Section XIII).

In practice, however, in addition to having access to a restricted set of accessible solutions due to the use of a single SLM, a second important constraint applies: the approximation $\sigma_{\vec r}\approx 0$ is not strictly satisfied (Supplementary Section IX). The general solutions given by Equation~\eqref{eq3ter} are therefore not only approximate but also difficult to implement. This is confirmed experimentally: adding a random binary $0/\pi$ phase pattern to the classical correction (Fig.~\ref{fig:Fig3}a), an exact solution of Equation~\eqref{eq3ter}, does not allow any image to be retrieved at the output (Figs.~\ref{fig:Fig3}b and c).

To deal with these limitations, we use an optimization approach to determine the non-classical solution $S'$.
A two-photon guide state is prepared at the input by removing both the object and the $f_0$ lens. This state closely approximates the maximally entangled state i.e. $ \int a^\dagger (\vec{r}) a^\dagger (-\vec{r}) \ket{0} d \vec{r}$ and, as shown in Figure~\ref{fig:Fig3}d, produces a sharp peak at the center of the correlation image when no medium is present.
After propagation through the complex medium, the peak becomes diffuse and significantly broadened (Fig.~\ref{fig:Fig3}e). To restore it, we use a partitioning algorithm on the SLM~\cite{vellekoop_phase_2008}, with the central correlation value serving as the optimization target. Since this procedure is time-consuming, we perform such an optimization numerically using a transmission matrix measured experimentally beforehand (see Supplementary Section~VIII). After more than $4000$ optimization steps (Fig.~\ref{fig:Fig3}g), a plateau is reached and the resulting SLM pattern (Fig.~\ref{fig:Fig2a}c) enables the recovery of a sharp peak in the output correlation image (Fig.~\ref{fig:Fig3}f).

By analogy with classical methods~\cite{katz_looking_2012,horstmeyer_guidestar-assisted_2015}, the maximally entangled state effectively acts as a guidestar in the correlation image, which the optimization process seeks to restore.
This process drives the system toward a non-trivial {transformation} $S'$. To gain insight into this solution, we subtract the classical phase mask (Fig.~\ref{fig:Fig2a}b) from the non-classical one (Fig.~\ref{fig:Fig2a}c). The resulting phase mask, shown in Fig.~\ref{fig:Fig3}h, is neither uniform nor limited to values of $0$ and $\pi$, but instead displays phase values distributed between $0$ and $2\pi$, with two distinct, broad peaks separated by approximately $\pi$ (Fig.~\ref{fig:Fig3}i).
A detailed comparison between the experimental solutions and those derived from Equation~\eqref{eq3ter} is provided in Supplementary Section X. Moreover, note that the use of a classical guidestar does not allow convergence toward non-classical solutions, as shown in Supplementary Section XI.

\section*{Demonstration in other complex media}

\begin{figure*}
    \centering
    \includegraphics[width=0.99\linewidth]{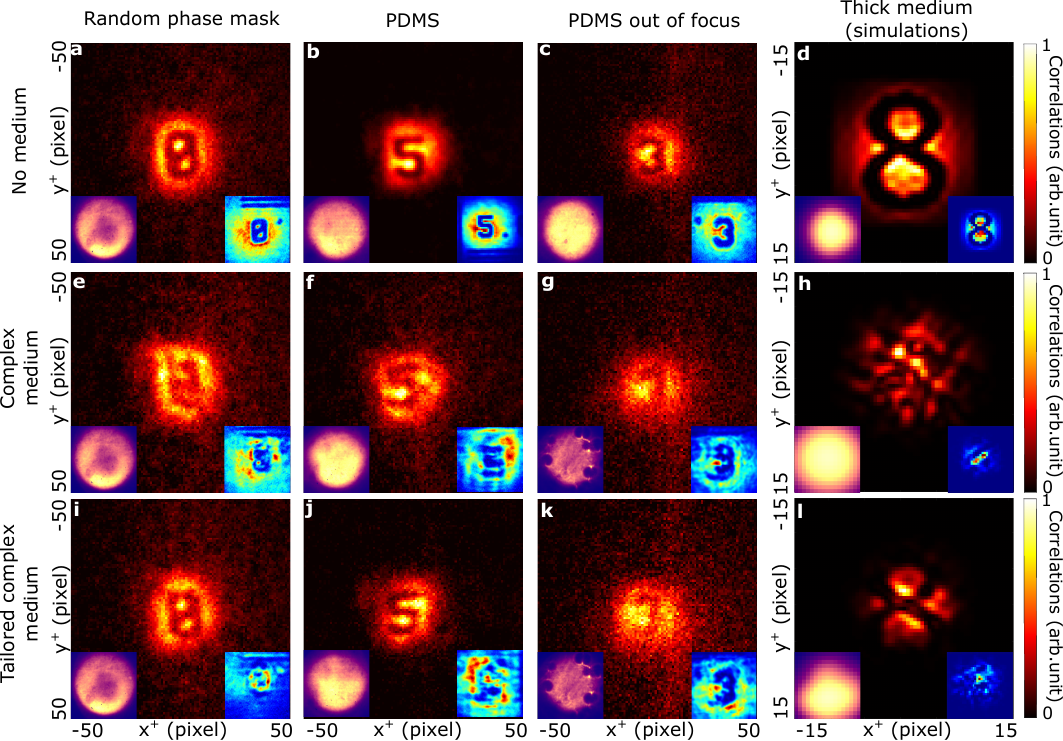}
    \caption{\textbf{Results for different objects and complex media.} Experimental \textbf{(a-c)} and simulated \textbf{(d)} correlation images obtained with different objects and no medium. 
    \textbf{e-h,} Correlation images obtained {with the complex medium and no tailoring. Complex media includes a random phase mask with controllable disorder displayed on the SLM \textbf{(e)}, a 1-cm-thick layer of polydimethylsiloxane (PDMS) positioned in a plane conjugate to the SLM \textbf{(f)} and out of plane \textbf{(g)}, and a simulated thick scattering medium \textbf{(h)}.}
    \textbf{i-l,} Correlation images obtained after tailoring the medium. Insets at the bottom left of each panel show intensity images measured under photon-pair illumination; bottom right insets show the corresponding intensity images acquired under classical illumination. Each image is normalized to its maximum value. See Methods for more details on the simulations.}
    \label{fig:Fig4}
\end{figure*}

To evaluate the versatility of our method, we applied it to various objects and different types of complex media. For example, Figures~\ref{fig:Fig4}a,e and i show the measured correlation and intensity images obtained through a random phase mask displayed on the SLM (instead of the physical parafilm layer), where the level of disorder can be precisely controlled. In this case as well, the system becomes transparent to quantum correlations while remaining opaque in classical intensity imaging. This effect holds for different levels of disorder, provided that the number of active SLM pixels remains larger than the correlation width of the applied phase mask (see Supplementary Section {XII}). 
To better approximate real imaging conditions, we also replaced the scattering layer with a 1-cm-thick layer of polydimethylsiloxane (PDMS), a material commonly used in microscopy to simulate low-order aberrations~\cite{kang_high-resolution_2017,cameron_adaptive_2024}. Whether placed in a plane conjugate to the SLM (Figs.~\ref{fig:Fig4}b, f, and j) or in a non-conjugate plane (Figs.~\ref{fig:Fig4}c, g, and k), our method remains effective, correcting aberrations in the correlation image but not in the classical intensity image.
For more complex and thicker media, experiments are challenging to perform due to the very low transmitted photon flux. However, numerical simulations using matrices that reproduce light scattering in thick media show that the approach still works (Figs.~\ref{fig:Fig4}d, h, and l), albeit over a reduced FOV, similar to that encountered in classical wavefront shaping {and ultimately constrained by the optical memory effect of the medium~\cite{katz_looking_2012,freund_memory_1988}}. The optimized phase masks obtained in these four cases are presented in Supplementary section {X}. Finally, extending our approach to dynamic media - more representative of realistic conditions - will require faster correlation imaging and optimization, achievable with improved cameras or algorithms~\cite{hogenbirk_intensified_2025,conkey_genetic_2012}.

\section*{Comparing classical and quantum image transmission}

The key advantage of our approach is its ability to selectively suppress information encoded in classical states by scattering it, while preserving information encoded in entangled states. To quantitatively assess the performance of this filtering, we analyze the image transmission fidelity obtained with the classical solution and with 50 distinct non-classical solutions, as shown in Figure~\ref{fig:Fig5}a. The non-classical solutions are obtained by repeating the numerical optimization process with different initial phase masks using the experimentally measured transmission matrix of Figure~\ref{fig:Fig2a}. While the classical solution yields a transmission fidelity exceeding $0.9$ for both classically and quantum-encoded images, programming a non-classical solution reduces the fidelity of classically encoded images to an average value of $0.25$, thereby confirming the strong filtering of classical information (see Methods).

Importantly, the non-classical solutions also scatter images encoded in classical correlations. Figures~\ref{fig:Fig5}b–d show simulated image transmissions using a non-classical solution for an image encoded in an entangled state (b), in the intensity of a classical state (c), and in the correlations of a separable mixed two-photon state (d). Only the image encoded in the entangled state is faithfully transmitted (see Supplementary Section~XI). This demonstrates that the filtering mechanism is inherently sensitive to entanglement, rather than to the mere presence of optical correlations.}

\begin{figure}
    \centering
\includegraphics[width=1\linewidth]{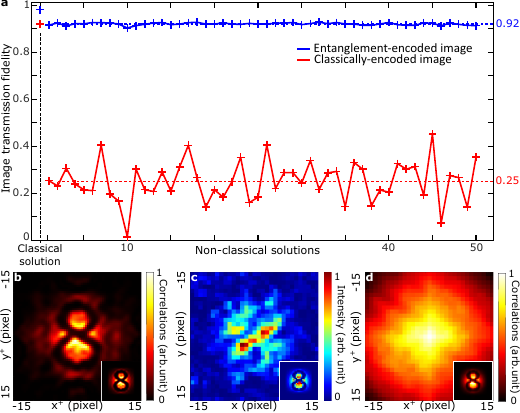}
\caption{\textbf{Quantitative evaluation of quantum-classical image filtering.}
\textbf{a,} Image transmission fidelity for the classical solution (abscissa 0) and for $50$ non-classical solutions obtained by numerical optimization. As detailed in Methods, the fidelity is computed using the structural image similarity (SSIM) metric between the transmitted image and the reference image obtained in the absence of the scattering medium. Images are generated through numerical simulations using a point source as the object and an experimentally measured transmission matrix. The average transmission fidelity over the $50$ non-classical patterns is $0.92\pm 0.0007$ for the entangled state (blue line) and $0.25\pm0.01$ for the classical encoding (red line). \textbf{b-d,} Simulated image transmission using a non-classical solution for an image encoded in the entangled state \textbf{(b)}, in the intensity of a classical state \textbf{(c)}, and in the correlations of a separable two-photon state \textbf{(d)}. Insets show the corresponding reference images without the scattering medium. Additional details are provided in Supplementary Sections~XI and XVI.}
    \label{fig:Fig5}
\end{figure}

\section*{Conclusion}

Our approach leverages a unique property of entangled two-photon states, with no classical counterpart: the preservation of photon correlations across multiple optical bases~\cite{spengler_entanglement_2012}. This fundamental property is widely exploited in protocols such as entanglement certification and state tomography~\cite{wootters_optimal_1989,Bavaresco2018}. In our work, we apply this concept in an imaging context involving two distinct spatial bases: (i) the imaging basis, corresponding to the input image, and (ii) the tailored scattering medium basis $S'$. The similarity between the correlation images measured in these two different bases validates the success of our method - a result that cannot be replicated with classically correlated photons, as demonstrated in Figures~\ref{fig:Fig5}b-d and Supplementary Section {XI}.

In this work, we restrict our demonstration to a specific class of input entangled states, simple scattering media, and limited wavefront control (a single SLM), which constrains the class of non-trivial transformations $S'$ {accessible in practice} (Eq.~\eqref{eq3ter}). 
However, our approach extends well beyond this minimal setting.
For instance, Supplementary Section~XIV shows that alternative encodings - such as engineering the nonlinear crystal geometry rather than modulating the pump~\cite{yesharim_direct_2023} - leads to a distinct class of non-trivial {transformations} for $S'$. 
More generally, we analytically demonstrate in Supplementary Section~XVII that, for arbitrary input entangled states and arbitrarily complex scattering media, non-classical solutions that enable faithful quantum-state transmission while classical information remains scrambled always exist. Such solutions can, in principle, be implemented using multi-plane wavefront-shaping techniques~\cite{kupianskyi_all-optically_2024}.

Ultimately, by demonstrating that optical disorder can be tailored into a selective quantum–classical filter, our work introduces a conceptual shift: complex media can be harnessed as active, programmable components for quantum technologies, rather than being treated merely as obstacles to invert. In the realm of secure communication, this physical discrimination provides a fundamental advantage. Because our approach is physically sensitive to the presence of entanglement - not just classical correlations - it establishes a hardware-level protection against spoofing or blinding attacks, going beyond existing schemes like quantum illumination~\cite{lloyd_enhanced_2008,gregory_imaging_2020,defienne_quantum_2019}.
Looking further ahead, our strategy could inspire breakthrough imaging techniques in complex environment, such as biological tissues. It indeed offers multiple accessible solutions (see Supplementary Section {XIII}) without requiring inversion of the scattering process, and benefits from multimode optimization, which can increase the isoplanatic patch size (i.e. FOV), albeit at the cost of reduced contrast.

\section*{Acknowledgments}

H.D. acknowledges funding from the ERC Starting Grant (No. SQIMIC-101039375) and the ANR (ANR-24-CE97-0001 and ANR-23-CE47-0014).

\section*{Author Contributions}
C.V. analyzed the data, designed and performed the experiments, with support from B.C. and R.G. C.V. and H.D. conceived the original ideal. All authors discussed the results and contributed to the manuscript. H.D supervised the project.

\section*{Competing interests}
The authors declare no competing interests.

\section*{Methods}
    
\noindent \textbf{Correlation image.} The correlation image $\Gamma^+$ is the sum-coordinate projection of the spatial second-order correlation function $G^{(2)}$. To measure it, we use the Tpx3Cam camera, which provides a time-tagged list of photon detection events for each acquisition~\cite{nomerotski_intensified_2023}. Among these events, genuine coincidences are identified as pairs $(x_i, y_i, t_i)$ and $(x_s, y_s, t_s)$ that satisfy $|t_i - t_s| < 6\,\text{ns}$. All other pairs are treated as accidental and discarded. By definition, $G^{(2)}(x_i, y_i, x_s, y_s)$ is the number of genuine coincidences between the positions $(x_i, y_i)$ and $(x_s, y_s)$, and $\Gamma^+(x_+, y_+)$ is the number of such coincidences for which the position sum satisfies $(x_i + x_s, y_i + y_s) = (x_+, y_+)$. Formally, the two quantities are related by: 
\begin{equation}
\label{eq4bis}
    \Gamma^+(\vec{r}_+) = \iint G^{(2)}(\vec{r}, \vec{r}_+ - \vec{r}) \, d\vec{r}.
\end{equation}
$\Gamma^+$ is interpreted as a probability map of the photon-pair barycenter. For example, in the case of near-perfect anti-correlations i.e. when a photon detected at $\vec{r}_i$ implies its twin is likely at $\vec{r}_s = -\vec{r}_i$ - as observed when the object and lens $f_0$ are removed - the barycenter remains close to the image center, a sharp peak appears at the center of the correlation image $\Gamma^+$. In our work, all correlation images were acquired using acquisition time ranging between $2$s and $30$s. Further details are provided in Supplementary Section II.\\
\\
\noindent \textbf{Matrix formalism}.
Assuming that the photon pairs occupy the same polarization and spectral mode, the two-photon state generated by the crystal can be written as:
\begin{equation}
\label{eq4}
|\psi \rangle = A \iint \psi^{\text{in}}({\vec{r_i}, \vec{r_s}})  a^\dagger_{\vec{r_i}} a^\dagger_{\vec{r_s}} | 0 \rangle d\vec{r_i} d\vec{r_s},
\end{equation}
where $A$ is a normalization constant, $\psi^{\text{in}}$ is the spatial two-photon wavefunction, and $\vec{r_i} = (x_i, y_i)$, $\vec{r_s} = (x_s, y_s)$ are the transverse spatial coordinates of the idler and signal photons, respectively, in the input plane.
Upon propagation through an arbitrary complex linear optical system, the two-photon wavefunction in the output plane becomes~\cite{abouraddy_entangled-photon_2002}:
\begin{equation}
\label{eq5}
\psi^{\text{out}}({\vec{r_i'}, \vec{r_s'}}) = \iint \psi^{\text{in}}({\vec{r_i}, \vec{r_s}}) s(\vec{r_i'}, \vec{r_i}) s(\vec{r_s'}, \vec{r_s}) d\vec{r_i} d\vec{r_s},
\end{equation}
where $\psi^{\text{out}}$ is the output wavefunction, $\vec{r_i'}$ and $\vec{r_s'}$ are the transverse positions in the output plane, and $s$ is the coherent point spread function (PSF) of the system. After discretization and ordering of the transverse positions, the wavefunctions $\psi^{\text{in}}$, $\psi^{\text{out}}$, and the PSF $s$ can be represented by matrices $\Psi^{\mathbf{in}}$, $\Psi^{\mathbf{out}}$, and $S$, respectively. Equation~\eqref{eq5} then becomes the matrix equation given in Equation~\eqref{eq2}. \\
\\
\noindent \textbf{Image encoding}.
The second-order spatial correlation $G^{(2)}$ in the input plane is obtained by taking the squared modulus of the two-photon field given in Equation~\eqref{eq3}:
\begin{equation}
    G^{(2)}(\vec{r_i},\vec{r_s}) = \left| t \left(\frac{\vec{r_i}+\vec{r_s}}{M} \right) \exp{\frac{-|\vec{r_i}-\vec{r_s}|^2 \pi^2 \sigma_{\vec{r}}^2 }{4 \lambda_p^2 f_1^2}} \right| ^2.
\end{equation}
The correlation image $\Gamma^+$ measured in the camera plane (output plane) without the scattering medium is thus obtained using Equation~\eqref{eq4bis}, accounting for the magnification $M'$ between the input and output planes:
\begin{equation}
    \Gamma^+(\vec{r_+}) \approx C \left| t \left(\frac{\vec{r_+}}{M'M} \right) \right| ^2,
\end{equation}
where $C$ is a constant. Further details are provided in Refs.~\cite{HidingImages,abouraddy_entangled-photon_2002} and in Supplementary Section III.\\
\\
\noindent \textbf{{Non-trivial {transformations} $S'$}}. {Combining equations~\eqref{eq3} and \eqref{eq5}, the two-photon field in the output plane can be simplified as:
\begin{equation}
\label{eq8}
    \psi^{\text{out}}({\vec{r_i'}, \vec{r_s'}}) = \int t(\vec{r_+}) H_{s'}(\vec{r_i'}, \vec{r_s'}, \vec{r_+}) d \vec{r_+} ,
\end{equation}
where $H$ is defined as:
\begin{eqnarray}
H_{s'}(\vec{r_i'}, \vec{r_s'}, \vec{r_+}) &=& \int s' \left(\vec{r_i'}, \frac{M \vec{r_+}}{2} +\vec{r_-} \right) \nonumber \\
&&s'\left (\vec{r_s'}, \frac{M \vec{r_+}}{2}-\vec{r_-} \right) \nonumber  \\
 \label{eq9} &&\exp{\frac{- |\vec{r_-}|^2 \pi^2 \sigma_{\vec{r}}^2 }{ \lambda_p^2 f_1^2}} d\vec{r_-},
\end{eqnarray}
where $s'$ is the coherent PSF associated with $S'$. For $S'$ to be a non-trivial transformation, $s'$ must simultaneously verify: 
\begin{equation}
s'(\vec{r'},\vec{r}) \neq \delta \left(\vec{r'} + \alpha \vec{r} \right),
\end{equation}
 and
\begin{equation}
\label{eq7}
H_{s'}(\vec{r_i'}, \vec{r_s'}, \vec{r_+}) = \delta \left( (\vec{r_i'}  \pm \vec{r_s'}) - \vec{r_+}\right),
\end{equation}
where $\alpha \in [-1,1]$ are arbitrary coefficient taking into account potential magnification. 
In the limit $\sigma_{\vec{r}} \rightarrow 0$, equation~\eqref{eq9} simplifies to:
\begin{equation}
H_{s'}(\vec{r_i'}, \vec{r_s'}, \vec{r_+}) = \int s'(\vec{r_i'}, M \vec{r_+}/2 + \vec{r_-}) \, s'(\vec{r_s'}, M \vec{r_+}/2 - \vec{r_-}) \, d\vec{r_-}.
\end{equation}
Substituting the general form of $s'$ from equation~\eqref{eq3ter} into the expression above yields:
\begin{equation}
H_{s'}(\vec{r_i'}, \vec{r_s'}, \vec{r_+}) = \delta \left( \vec{r_i'}+\vec{r_s'}-M \vec{r_+} \right),
\end{equation}
which, when inserted into equation~\eqref{eq8}, shows that $\psi^{\text{out}}(\vec{r_i'}, \vec{r_s'}) = t((\vec{r_i'} + \vec{r_s'})/M)$; that is, the object is perfectly restored at the output. See Supplementary Sections III and {XIV} for more details.} \\

\noindent \textbf{Numerical simulations}. All numerical simulations were performed using MATLAB. The input states were propagated via matrix multiplication following the formalism described in Equation~\eqref{eq2}. To optimize computation time, the simulations were carried out on a discretized transverse space of size less than $51 \times 51$ pixels, resulting in matrices of size $2601 \times 2601$.
Unless otherwise specified, both the input state $\Psi^{\text{in}}$ and the propagation matrix $S$ used in the simulations were experimentally characterized beforehand to closely match the experimental conditions. 
Further details on the numerical simulations are provided in Supplementary Sections IV.\\
\\ 
\noindent \textbf{Optimization process}. The optimization process is used to determine the SLM phase mask that implements the optical transformation $S'$. It is carried out by numerically simulating the experiment, following an initial experimental characterization phase.\\
First, the input state - corresponding to the state produced at the output of the crystal without the object and lens $f_0$ - is characterized experimentally from spatial correlation measurements in both the input and SLM planes, following the methods described in Refs.~\cite{Edgar2012,defienne_general_2018,courme_quantifying_2023}. The state is then modeled as a Double-Gaussian distribution~\cite{Fedorov2009} and discretized into two matrices $\Psi^{\text{in}}$ and $\Psi^{\text{SLM}}$, the former describing the state in the input plane and the latter in the SLM plane.
Second, the matrix $S_m$ that links the SLM plane to the output plane (i.e. including the scattering medium) is experimentally measured. For this, the SLM is discretized into $32 \times 32$ macropixels, and the matrix is obtained using a classical coherent light source following the method described in Ref.~\cite{gigan_matrice_2010}. This measurement enables us to construct the full imaging system matrix i.e. the matrix that links the input to the output plane, denoted $S'$. It is given by the product $S' = S_m D_{\text{SLM}} F_{f_1} $, where $D_{\text{SLM}}$ is a diagonal phase matrix representing the action of the SLM, and $F_{f_1}$ is a discrete Fourier transform performed by the lens $f_1$.\\
Following this characterization phase, the optimization consists of numerically propagating the input state $\Psi^{\text{in}}$ through the optical system $S'$ using Equation~\eqref{eq2}, for various phase masks displayed on the virtual SLM - that is, by modifying the phase terms in $D_{\text{SLM}}$ - in order to optimize a target in the output correlation image. In practice, to minimize numerical errors introduced by the discrete Fourier transform $F_{f_1}$, we simulate the propagation of the experimentally measured state in the SLM plane, $\Psi^{\text{SLM}}$, through the system described by $S_m$. In this case, the propagation is given by: $\Psi^{\text{out}} = S_m D_\text{SLM} \Psi^{\text{SLM}} (S_m D_\text{SLM}) ^t$.
At each optimization step, half of the SLM macropixels - randomly selected - are modulated with a phase delay swept from $0$ to $2\pi$ in {7} steps. This is known as a partitioning algorithm~\cite{vellekoop_phase_2008}. For each phase value, the numerical propagation is performed, and the central value of the output correlation image, noted $\Gamma^+_0$, is recorded. As shown in Ref.~\cite{courme2025nonclassicaloptimizationcomplexmedia}, these data are fitted with the function $a \cos(\theta + \theta_a) + b \cos(2 \theta + \theta_b) + c$
where $a$, $b$, $c$, $\theta_a$, and $\theta_b$ are fitting parameters, and $\theta$ is the applied phase shift. The phase shift $\theta_{opt}$ that maximizes this function is then applied to all pixels in the selected group. The procedure is repeated iteratively with new randomly selected subsets of pixels.\\
At each step, the central peak value of the correlation image increases. As shown in Figure~\ref{fig:Fig4}d, approximately $4000$ optimization steps are required to reach a plateau. At the end of the optimization, a phase pattern $D_{\text{SLM}}$ that maximizes the correlation peak is obtained, thereby realizing the transformation $S'$. This phase pattern is then programmed onto the SLM in the real experiment, assuming that the experimental conditions remain stable over the tens of minutes required for performing the numerical optimization.
Further details on the optimization procedure, as well as the experimental characterization of the input state and scattering matrix, are provided in Supplementary Sections V, VI and VII.\\
\\
\noindent \textbf{Image transmission fidelity}. The image transmission fidelity values shown in Fig.~\ref{fig:Fig5}a are obtained by computing the structural similarity index (SSIM) between an output image (classical intensity or quantum correlation) after propagation through the system $S'$, composed of the scattering medium and the SLM, and a reference image acquired without the scattering medium. The output images are obtained via numerical simulations using a point-source as the object and an experimentally measured transmission matrix of the scattering medium, the same as the one used to generate the results in Fig.~\ref{fig:Fig2b}. The transmission fidelity is defined by normalizing the SSIM as $(\mathrm{SSIM}-\mathrm{SSIM}_0)/(1-\mathrm{SSIM}_0)$, where $\mathrm{SSIM}_0$ is the mean SSIM between the reference image and images obtained through the scattering medium using random phase patterns on the SLM. Additional details are provided in Supplementary Section~XVI.

\section*{Data availability}
The data that support the findings of this study are available in the figshare repository under the identifier https://doi.org/10.6084/m9.figshare.7730219. Any additional data are available from the corresponding author upon reasonable request.

\end{document}


\preprint{APS/123-QED}

\tableofcontents

\section{Details on the experimental setups}
\label{setupconf}

\subsection{Experimental devices}

\noindent \textbf{Sources:} A superluminescent diode (SLED) centered at $810$nm with $10$nm bandwidth is used to perform classical intensity imaging and scattering matrix measurements. It is superimposed on the pump laser, a 100mW-power continuous-wave vertically polarised diode laser at 402nm (Coherent). To produce photon pairs, this laser is illuminating a $1 \times 5 \times 5$mm type-I $\beta$-barium borate (BBO) crystal (Newlight Photonics). Photons pairs at $810$nm are generated by Spontaneous Parametric Down-Conversion (SPDC) and near-collinear phase matching of photons at the output is ensured by slightly tilting the crystal around the horizontal axis.\\
\\
\noindent \textbf{Cameras:} A CS165MU - Zelux (Thorlabs) Complementary Metal-Oxide-Semiconductor (CMOS) camera is used to do some of the classical intensity imaging (Figures 3e,f,g,h and Figure 4b of the main document). An electron multiplied charge-coupled device (EMCCD) (iXon Life 897, Andor) is used is perform all the other classical intensity images (Figure 5 of the main document). A Tpx3Cam camera (Amsterdam Scientific Instrument) is used to detect the photon pairs. This event-based camera has pixels of size $55 \times 55 \mu$m$^2$ and a temporal resolution of $1.56$ns. The Tpx3cam camera is an advanced intensified hybrid CMOS sensor. To achieve single-photon sensitivity, incoming photons are first amplified using a photon intensifier (Photonis) mounted in front of the sensor. The resulting bursts of photons are processed by a centroid algorithm to identify the time and position of the incident photons. Consequently, the camera returns a list of discrete events $(x, y, t)$ - each corresponding to a single photon.\\
\\
\noindent \textbf{SLM:} The SLM is a phase only modulator (Holoeye Pluto-2-NIR-015) with $1920 \times 1080$ pixels and a $8\mu$m pixel pitch.

\subsection{Experimental configurations}

\noindent \textbf{Configuration for correlation images measurement:} The correlation image is created by illuminating the non-linear crystal with the Fourier transform of the desired object. The collimated laser beam is magnified by a telescope $f'_1 = 50\text{mm}- f'_2=150\text{mm}$ before illuminating the object, which is located in the object plane of a lens $f_0$ of focal length $f_0=200$mm. {For the data shown in Figure 5 of the main document, the laser beam is not magnified, so $f'_1$ and $f'_2$ are absent, and $f_0=75$mm.}
The lens $f_1 = 35\,\text{mm}$ Fourier images the crystal onto the input plane of the system.
The lens $f_2$ is composed of three lenses in a confocal configuration: $60\text{mm}-50\text{mm}-125\text{mm}$.
A telescope with focal lengths $f_3=200\text{mm}$-$f_4=75\text{mm}$ images the SLM plane onto the plane where the scattering medium is positioned.
The output surface of the scattering medium is then Fourier imaged onto the output plane by the lens $f_5$. For the data shown in Figures 3 and 4 of the main document, $f_5 = 125\,\text{mm}$, and for Figure 5, $f_5 = 100\,\text{mm}$.
The lens $f_c$, which represents a unit-magnification imaging system mapping the output plane onto the Tpx3Cam sensor, is shown in Figure 2a for clarity only. In practice, it is not present, and the Tpx3Cam is placed directly at the output plane i.e. at the focal plane of lens $f_5$.\\
\textit{Magnifications: }The magnification $M$, defined as $M = 2f_1 / f_0$, is $M = 0.93$.
The magnification $M'$ between the input and output planes is $M' = 0.83$ for the results in Figures 3 and 4, and $M' = 0.66$ for Figure 5.
The magnification from the crystal plane to the SLM plane is $M''=4.3$.
The magnification from the crystal plane to the scattering medium plane is $M'''=1.6$.
The effective focal length from the crystal plane to the output plane is $78$mm for the results shown in Figures 3 and 4, and $62.5$mm for Figure 5.\\
\\
\noindent \textbf{Configuration for classical intensity image measurement:} For intensity images acquired with the Zelux camera (Figure 3 of the manuscript), the lens $f_i$ is not present, as the camera is placed directly at the output plane, and $f_5 = 150\,\text{mm}$. For intensity images acquired with the EMCCD camera (Figure 5 of the manuscript), $f_5 = 125\,\text{mm}$, and the lens $f_i$ consists of two lenses in a telescope configuration: $150\,\text{mm}$–$150\,\text{mm}$.

\section{Measuring correlation images with the tpx3cam }

\noindent \textbf{Spatial $G^{(2)}$ measurement:} Detection of the photon pairs is carried out with a Tpx3cam camera. The camera returns a time-tagged list of events ($x$, $y$, $t$), with $r=(x, y)$ denoting the position of the detected photon and $t$ its detection time. Two events $(x_1,y_1,t_1)$ and $(x_2,y_2,t_2)$ are identified as a photon pair if they are detected within the same $6$ ns temporal window, \textit{i.e.} if $\lvert{t_1}-t_2 \rvert<6$ ns. Unlike traditional frame-based cameras (e.g. EMCCD) which require frame accumulation to statistically estimate correlations~\cite{defienne_general_2018}, the event-based Tpx3cam allows for real-time identification of pixel-level coincidences. This enables to build the spatial second-order correlation function $G^{(2)}(\vec{r_1},\vec{r_2})$ without frame averaging. All data presented in the main document were acquired using acquisition times ranging from $2$ to $30$ seconds. More details about the Tpx3cam operation can be found in Ref.~\cite{nomerotski_intensified_2023}.\\
\\
\noindent \textbf{Sum-coordinate projection:} The correlation image, noted $\Gamma^+$, corresponds to the projection of $G^{(2)}$ along the sum-coordinate axis. It is defined as follows (Equation (6) in the manuscript):
\begin{equation}
\label{eqSM1}
    \Gamma^+(\vec{r}_+) = \iint G^{(2)}(\vec{r}, \vec{r}_+ - \vec{r}) \, d\vec{r}.
\end{equation}
{To better understand its meaning, one can consider the following change of variables: $\vec{r}_+ = (\vec{r}_s + \vec{r}_i)/2$ and $\vec{r}_- = (\vec{r}_s - \vec{r}_i)/2$. Then, one can define the function $G_\pm^{(2)}$ such that $G_\pm^{(2)}(\vec{r}_+, \vec{r}_-) = G^{(2)}(\vec{r}_i, \vec{r}_s)$. With this definition, Equation~\eqref{eqSM1} can be rewritten as:
\begin{equation}
    \Gamma^+(\vec{r}_+) = \iint G^{(2)}(\vec{r}, \vec{r}_+ - \vec{r}) \, d\vec{r} = \iint G_\pm^{(2)}(\vec{r_+}, \vec{r_-}) \, d\vec{r_-}.
\end{equation}
From the above equation, it follows that the sum-coordinate projection $\Gamma^+$ is obtained by integrating the second-order correlation function along the diagonal $\vec{r}_- = \vec{r}_s - \vec{r}_i$, thereby projecting it onto the sum-coordinate $\vec{r}_+ = \vec{r}_s + \vec{r}_i$.}

\begin{figure}
    \centering
    \includegraphics[width=0.8\linewidth]{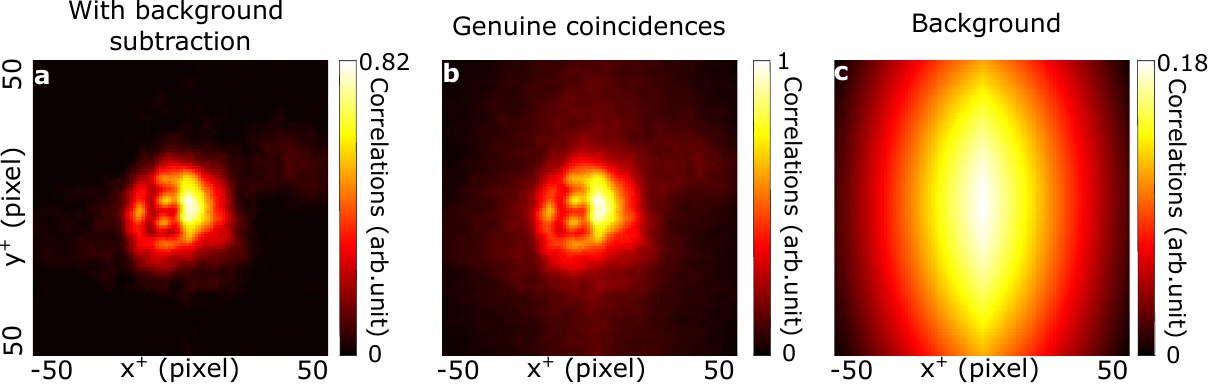}
    \caption{\textbf{Impact of accidental coincidences}. \textbf{a}, Correlation image shown in Figure 2b of the main text, obtained by subtracting an image measured via direct coincidence counting with the camera \textbf{(b)} from an image containing only accidental coincidences \textbf{(c)}, estimated from the single detection events.
}
    \label{fig:SM_background}
\end{figure}

In contrast to frame-based cameras, which often require background subtraction to eliminate accidental coincidences~\cite{defienne_general_2018} (i.e. uncorrelated photons or noise), the high temporal resolution of the Tpx3cam camera can significantly reduce such artifacts, depending on the chosen coincidence window~\cite{guitter_accidental_2025}. However, because disordered media can introduce losses and noise, we still perform background subtraction to enhance the visibility of the correlation image. Figure~\ref{fig:SM_background}a shows the correlation image of a $8$ corresponding to the one shown in Figure 3d of the manuscript. Figure~\ref{fig:SM_background}b shows the image obtained from genuine photon coincidence detection with the camera (without accidental subtraction), while Figure~\ref{fig:SM_background}c shows the image containing only accidental coincidences, estimated directly from the single detection events. Subtracting the image in Figure~\ref{fig:SM_background}c from that in Figure~\ref{fig:SM_background}b yields the final correlation image shown in Figure~\ref{fig:SM_background}a. 

\section{Details on the quantum-image encoding scheme and demonstration of Equations (3) and (10) of the manuscript}
\label{math3and10}
The correlation image with the object is formed following the method described in Ref.~\cite{HidingImages}. In this section, we provide further details about the two-photon state thus created and demonstrate Equations (3) and (10) of the manuscript. To do so, we follow the reasoning and general calculations presented in Section 3 of Ref.~\cite{Schneeloch_2016}.
We begin by considering the following reasonable assumptions:
(i) The pump is nearly monochromatic;
(ii) The spatial spectra of the pump and photon pairs are dominated by their longitudinal components (paraxial regime);
(iii) The pump is bright enough to be treated classically, and its attenuation due to downconversion is negligible;
(iv) The crystal dimensions are much larger than the wavelengths of the pump and photon pairs;
(v) A frequency filter is placed before the camera, so that only degenerate downconversion events ($\lambda_s = \lambda_i$) are considered.

\subsection{Calculation of the two-photon wavefunction in momentum space at the crystal plane $\tilde{\psi}^{\text{c}}$}

Under these assumptions, the two-photon wavefunction in momentum space at the crystal plane reads 
\begin{equation}
\left| \psi^{\text{c}} \right\rangle = \iint d\mathbf{k}_s \, d\mathbf{k}_i \, \tilde{\psi}^{\text{c}}(\mathbf{k}_s, \mathbf{k}_i) \, \hat{a}_s^\dagger(\mathbf{k}_s) \, \hat{a}_i^\dagger(\mathbf{k}_i) \, \left| 0, 0 \right\rangle,
\end{equation}

where $\mathbf{k}_s$ and $\mathbf{k}_i$ are the wave vectors of the signal and idler photons, respectively, $\hat{a}_s^\dagger(\vec{k}_s)$ and $\hat{a}_i^\dagger(\vec{k}_i)$ are their corresponding creation operators, and $\tilde{\psi}^{\text{c}}$ is the two-photon wavefunction in momentum space $(\vec{k_s},\vec{k_i})$ at the crystal plane. It can be expressed as:
\begin{equation}
\label{equFontionK}
\tilde{\psi}^{\text{c}}(\vec{k}_s, \vec{k}_i) = \mathcal{N}\, \text{sinc}\left( \frac{\Delta k_z L}{2} \right) \mathcal{F}\{E^p\}(\vec{k}_s+\vec{k}_i),
\end{equation}
where $\mathcal{N}$ is a normalization factor, $E^p$ is the transverse profile of the pump beam at the crystal plane, $\mathcal{F}$ is a Fourier transform, $L$ is the length of the crystal, $\Delta k_z = k_{pz} - k_{sz} - k_{iz}$ is the longitudinal phase mismatch, with $k_{pz}$, $k_{sz}$, and $k_{iz}$ being the longitudinal components of the pump, signal, and idler wave vectors, respectively. Equation~\eqref{equFontionK} corresponds to Equation (21) in Ref.~\cite{Schneeloch_2016}. 
Using the assumption that the angles between the signal and idler momentum vectors and the pump momentum are small (paraxial regime), one can further simplify $\Delta k_z \approx -\frac{ \lambda_p |\vec{k}_s - \vec{k}_i|^2 }{4\pi}$~\cite{Schneeloch_2016}, where $\lambda_p = 402$nm is the pump wavelength in the crystal. The two-photon wavefunction then becomes:
\begin{equation}
\label{equFontionK2}
\tilde{\psi}^{\text{c}}(\vec{k}_s, \vec{k}_i) = \mathcal{N}\, \text{sinc}\left( -\frac{ L \lambda_p }{4  \pi } |\vec{k}_s-\vec{k}_i|^2 \right) \mathcal{F}\{E^p\}(\vec{k}_s+\vec{k}_i).
\end{equation}
Then, we make an additional assumption by approximating the \text{sinc} function with a Gaussian function, as detailed in Ref.~\cite{Schneeloch_2016}. This approximation closely matches experimental results and allows for simpler manipulation of the analytical expressions. Mathematically, the approximation used is the following:
\begin{equation}
    \frac{3}{4} \sqrt{\frac{a}{\pi}} \text{sinc} \left( a |\vec{k}|^2 \right) \approx \frac{1}{2 \pi b} \exp{- \frac{|\vec{k}|^2}{2 b^2}},
\end{equation}
where $b = \sqrt{\frac{3}{4a}}$. When applied to Equation~\eqref{equFontionK2}, it leads to:
\begin{equation}
    \text{sinc}\left( -\frac{ L \lambda_p }{4  \pi } |\vec{k}_s-\vec{k}_i|^2 \right) \approx C \exp{ -\frac{\sigma_{\vec{r}}^2}{4}\lvert \mathbf{k_i}-\mathbf{k_s}\rvert^2 },
\end{equation}
where $C$ is a constant and $\sigma_{\vec{r}} = \sqrt{\frac{2 L \lambda_p}{3 \pi}}$ is the position correlation width. The two-photon wavefunction in momentum space at the crystal plane then simplifies as 
\begin{equation}
\label{equFontionK3}
\tilde{\psi}^{\text{c}}(\vec{k}_s, \vec{k}_i) = \mathcal{N}\, \exp{ -\frac{\sigma_{\vec{r}}^2}{4}\lvert \mathbf{k_i}-\mathbf{k_s}\rvert^2 } \mathcal{F}\{E^p\}(\vec{k}_s+\vec{k}_i).
\end{equation}

\subsection{Calculation of the pump spectrum $\mathcal{F}\{E^p\}$}

In our experimental setup, the pump spectrum $\mathcal{F}\{E^p\}$ is determined by the object $t$ and the lens $f_0$ placed before the crystal. Since the lens performs an exact Fourier transform and assuming uniform illumination of the object by the pump, the pump spectrum matches the shape of the object, up to the scaling factor imposed by the Fourier transform of the lens:
\begin{equation}
\label{fieldEp}
    \mathcal{F}\{E^p\}(\vec{k}) = t \left ( \frac{\lambda_p f_0}{ 2 \pi}\vec{k} \right ),
\end{equation}
where $f_0$ is the focal length of the lens between the object and the crystal.

\subsection{Two-photon wavefunction in position space at the input plane $\psi^{\text{in}}$}

In our experimental configuration, the input plane corresponds to a Fourier plane of the crystal. To calculate the wavefunction at the input plane, denoted $\psi^{\text{in}}$, we substitute $\vec{k}$ in Equation~\eqref{equFontionK3} with $\frac{2 \pi \vec{r}}{\lambda f_1}$, where $\lambda = 804$ nm is the photon pair wavelength, $f_1$ is the focal length of the lens performing the Fourier transform between the crystal and the input plane, and $\vec{r}$ is the transverse spatial position in the input plane, yielding:
\begin{equation}
\label{equFontionK4}
\psi^{\text{in}}(\vec{r}_s, \vec{r}_i) = \mathcal{N}\, \exp{ -\frac{\pi^2 \sigma_{\vec{r}}^2}{4 \lambda_p^2 f_1^2}\lvert \mathbf{r_i}-\mathbf{r_s}\rvert^2 } \mathcal{F}\{E^p\} \left( \frac{2 \pi}{\lambda f_1}(\vec{r}_s+\vec{r}_i) \right),
\end{equation}
where we also used the fact that the photon are degenerated in frequency $2 \lambda_p=\lambda$. Finally, substituting Equation~\eqref{fieldEp} into Equation~\eqref{equFontionK4} yields Equation (3) of the manuscript:
\begin{equation}
\label{eq3manu}
\psi^{\text{in}}({\vec{r_s},\vec{r_i}}) = {t} \left(\frac{\vec{r_i}+\vec{r_s}}{M} \right) \exp{\frac{-|\vec{r_i}-\vec{r_s}|^2 \pi^2 \sigma_{\vec{r}}^2 }{4 \lambda_p^2 f_1^2}},
\end{equation}
where $M=\frac{2 f_1}{f_0}$.

\subsection{Visualization through the sum-coordinate projection $\Gamma^+$ and demonstration of Equation (10) of the manuscript}

\begin{figure}
    \centering
    \includegraphics[width=0.5\linewidth]{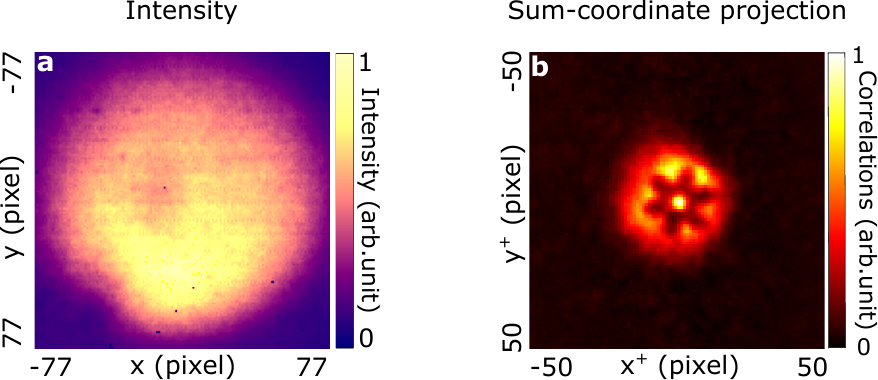}
    \caption{\textbf{Encoding images in quantum correlations}. \textbf{a}, Direct intensity on the camera that bears no trace of the object that we detect in \textbf{b}, the correlation image, showing a flower. Photon pairs are detected with the Tpx3cam camera. Acquisition time is $5$ seconds.}
    \label{fig:illustrationHiding}
\end{figure}

To visualize the encoded object, we measure the correlation image by projecting the measured second-order correlation function $G^{(2)}$ onto the sum-coordinate axis. Combining Equation~\eqref{eq3manu} with Equation~\eqref{eqSM1}, we find:
\begin{eqnarray}
\Gamma^+(\vec{r}_+) &=& \iint G^{(2)}(\vec{r}, \vec{r}_+ - \vec{r}) \, d\vec{r} \\
&=& \iint |\psi^{\text{in}}(\vec{r}, \vec{r}_+ - \vec{r})|^2 \, d\vec{r} \\
&=& \iint \left | {t} \left(\frac{\vec{r_+}}{M} \right) \right|^2 \exp{\frac{-|\vec{r_i}-\vec{r_s}|^2 \pi^2 \sigma_{\vec{r}}^2 }{4 \lambda_p^2 f_1^2}} \, d\vec{r} \\
&=& C \left | {t} \left(\frac{\vec{r_+}}{M} \right) \right|^2,
\end{eqnarray}
where $C$ is a constant. $\Gamma^+(\vec{r}_+)$ given by the above equation corresponds to the correlation image that would be measured if the camera were placed exactly in the input plane. When the camera is positioned in a conjugate image plane, as in our experiments when the scattering medium is removed, the magnification $M'$ between the input and output planes must also be taken into account, resulting in:

\begin{equation}
    \Gamma^+(\vec{r}_+) = C \left | {t} \left(\frac{ \vec{r_+}}{M'M} \right) \right|^2,
\end{equation}
which corresponds to Equation (10) of the manuscript. Figure~\ref{fig:illustrationHiding} illustrates the method. The intensity image presented in Figure~\ref{fig:illustrationHiding}a shows the incoherent SPDC beam that bears no trace of the object, while the correlation image shown in Figure~\ref{fig:illustrationHiding}b reveals the flower-shaped object. These images were acquired with the Tpx3cam camera over a $5$-second acquisition.

\section{General details on the numerical simulations}
\label{generaldetails}
\begin{figure}
    \centering
    \includegraphics[width=0.9\linewidth]{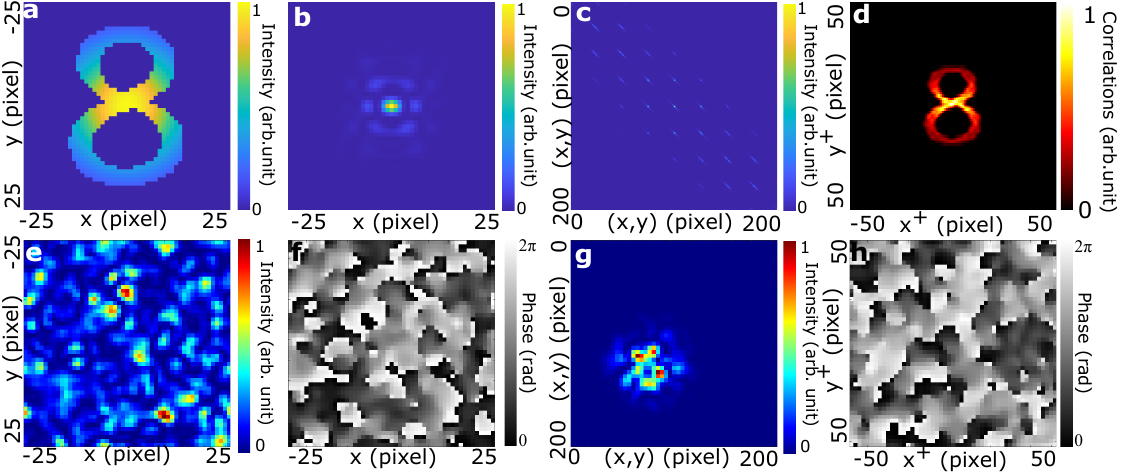}
    \caption{\textbf{Examples of simulations}. \textbf{a}, Classical intensity image of an object illuminated by a Gaussian beam. \textbf{b}, Insistency of the pump optical field in the crystal plane (i.e. Fourier transform of the object). \textbf{c}, Zoom into the central region of the squared modulus of the two-photon matrix in the crystal plane. \textbf{d}, Correlation image in the input plane. \textbf{e and f}, Amplitude and phase component of a speckle field, with correlation length $l_s=3$pixels, used to model the matrix $S_0$ of a thin scattering medium. \textbf{e and f}, Amplitude and phase component of the speckle field correponding to one column of a scattering matrix $S_0$ modeling a think scattering medium, defined by $l_s=3$pixels and $\sigma_s=8$pixels. }
    \label{fig:SM_simu}
\end{figure}

Unless otherwise specified, the simulations presented in the main text follow the method described in this section and are implemented using matrix operations in \textsc{Matlab}. Each simulation is divided into three parts: (i) pump shaping, (ii) two-photon field generation at the crystal plane, and (iii) propagation through the optical system i.e. from the crystal plane to any optical plane in the system.\\
\\
\noindent \textbf{(i) Pump shaping:} In this part, we simulate the spatial shape of the pump in the crystal plane. For that, the object (e.g. binary digit 8) is represented as a \( 51 \times 51 \) binary matrix corresponding to a physical area of \( 1 \times 1 \,\text{mm}^2 \). This object is illuminated by a Gaussian beam with a waist of \( 0.6 \,\text{mm} \), matching experimental measurements of the collimated laser beam. Figure~\ref{fig:SM_simu}a shows the resulting intensity image in the object plane, taking the form of a $51 \times 51$ pixel matrix in \textsc{Matlab}. The effect of a lens $f_0$ (between the object and the crystal) is simulated via a discrete Fourier transform over a grid of size \( 2601 \times 2601 \), yielding the amplitude and phase of the field at the crystal plane. Figure~\ref{fig:SM_simu}b shows the intensity image in the crystal plane. In \textsc{Matlab}, the pump field in the crystal plane, noted $E^p$, take the form of a $2601\times1$ complex vector.\\
\\
\noindent \textbf{(ii) Two-photon field generation:} To simulate the two-photon field $\psi^{\text{c}}$ generated at the crystal by a known pump field $E^p$, we use the following equation, as detailed in Ref.~\cite{Schneeloch_2016} (see also Section~~\ref{math3and10}):
\begin{equation}
\label{eq3manu2}
\psi^{\text{c}}({\vec{r_s},\vec{r_i}}) = E^p \left({\vec{r_i}+\vec{r_s}} \right) \exp{\frac{-|\vec{r_i}-\vec{r_s}|^2 }{\sigma_{\vec{r}}^2}},
\end{equation}
where $\vec{r_i}$ and $\vec{r_s}$ are the transverse spatial coordinate of the idler and signal photons in the crystal plane and $\sigma_{\vec{r}}$ is the position correlation width. The two-photon input field is simulated over a \( 1 \,\text{mm} \times 1 \,\text{mm} \) region, sampled with $51$ points along both \( x \) and \( y \) directions. The width $\sigma_{\vec{r}}$ depends only on the crystal thickness. It is characterized experimentally in Section~\ref{charactState} and set to {$13 \, \mu$m}.  In  \textsc{Matlab}, the two-photon field at the crystal plane is represented by a complex matrix \( \Psi^{\text{c}}\) of size \({2601 \times 2601} \). Figure~\ref{fig:SM_simu}b shows the squared absolute value of the two-photon field in the crystal plane.\\
\\
\noindent \textbf{(iii) Propagation through the optical system:} In general, the two-photon field is propagated through the optical system via matrix multiplication (Equation (2) of the manuscript), which is a discrete version of Equation (8) of the manuscript:
    \begin{equation}
        \Psi^{\text{out}} = S' \Psi^{\text{in}} (S')^t,
    \end{equation}
where $S'$ is the matrix linking the input to the output plane (encompassing the SLM and the scattering medium), and $\Psi^{\text{in}}$ and $\Psi^{\text{out}}$ are the two-photon matrices in the input and output planes, respectively. 
In our simulations, however, the system often needs to be decomposed into additional steps, in particular to model the SLM and the scattering medium. Starting from the crystal plane, the propagation can for example be written as follows:
    \begin{equation}
        \Psi^{\text{out}} =  F_{f_5} S_0 F_{f_4} F_{f_3} D_{\text{SLM}} F_{f_2} F_{f_1} \Psi^{\text{c}} (F_{f_5} S_0 F_{f_4} F_{f_3} D_{\text{SLM}} F_{f_2} F_{f_1})^t,
    \end{equation}
where $F_f$ denote the propagation matrix for a 2$f$-configuration (i.e free-space propagation over a distance $f$, followed by a lens of focal length $f$, and another free-space propagation over distance $f$). $D_{\text{SLM}}$ is the transfer matrix of the SLM, linking its input and output surfaces. In our simulations, it is modeled as a diagonal matrix containing $2601$ pure phase terms, corresponding to the $51 \times 51$ controllable pixels of the SLM screen. $S_0$ is the transfer matrix of the scattering medium, linking its input and output surfaces. Note that this matrix is different from the matrix $S$, which connects the input and output planes of the entire system and therefore includes the scattering medium as well as the surrounding lenses:
\begin{equation}
    S=F_{f_5} S_0 F_{f_4} F_{f_3} F_{f_2}.
\end{equation}
Following the same reasoning, the matrix $S'$, which corresponds to the tailored scattering medium, can be decomposed in the following general form:
\begin{equation}
    S'=F_{f_5} S_0 F_{f_4} F_{f_3} D_{\text{SLM}} F_{f_2}.
\end{equation}
In addition, we also define $S_m$ which is the matrix linking the SLM plane to the output plane,
\begin{equation}
    S_m=F_{f_5} S_0 F_{f_4} F_{f_3},
\end{equation}
and we also have the following simple link between the two-photon fields in the crystal and input planes:
\begin{equation}
\label{inputplane}
  \Psi^{\text{in}} = F_{f_1} \Psi^{\text{c}} F_{f_1}^t.
\end{equation}

In our simulations, depending on the objective, we adapt the system and model its various components (SLM, scattering medium, ...) based on known parameters. We then perform the corresponding matrix multiplications, starting from the two-photon field at the crystal plane, to compute the two-photon field at any plane within the system. Once the two-photon field is known in a given plane, we project its squared absolute value along the sum-coordinate axis to obtain the correlation image. {For example, Figure~\ref{fig:SM_simu}c shows a $200\times200$ submatrix of the simulated two-photon matrix correlation image in the crystal plane (absolute value) and Figure~\ref{fig:SM_simu}d shows the simulated correlation image in the input plane (i.e. using Equation~\eqref{inputplane}).}\\
\\
\noindent \textbf{Propagation of the classical field:} To simulate the intensity image of the object under classical illumination, we use the conventional matrix formalism from Ref.~\cite{gigan_matrice_2010} and captured by Equation (1) of the manuscript. To simulate the full system, we start from Equation (1) and decompose the optical setup into as many parts as needed:
    \begin{eqnarray}
        E^{\text{out}} &=&  S' E^{\textbf{in}} \\
        E^{\text{out}} &=&  S' F_{f_1} F_{f_0} E^{\textbf{obj}} \\
        E^{\text{out}} &=&  F_{f_5} S_0 F_{f_4} F_{f_3} D_{\text{SLM}} F_{f_2} F_{f_1} F_{f_0} E^{\textbf{obj}}.
    \end{eqnarray}
As in the case of photon pairs, the multiplications needed in each simulations are adapted according to the known parameters and the objective.\\
\\
\noindent \textbf{Modeling thin and thick scattering media:} In our simulations, we simulate the scattering medium using three different type of scattering matrices:\\
\\
\textit{Experimentally measured matrix $S_m$:} In some simulations, including those involved in the  the numerical optimization detailed in Section~\ref{Optimpro}, we use an experimentally measured transmission matrix $S_m$, linking the SLM plane to the output plane. The measurement of this matrix is detailed in Section~\ref{matrixmeasure}. When using this matrix, simulations are carried out using the following matrix propagation equation:  
    \begin{equation}
        \Psi^{\text{out}} =  S_m D_{\text{SLM}} F_{f_2} F_{f_1} \Psi^{\text{c}} (S_m D_{\text{SLM}} F_{f_2} F_{f_1})^t.
    \end{equation} \\
\\
\textit{Scattering matrix mimicking thin and thick scattering media $S_0$: } In the other simulations, we model the propagation through thin and thick scattering media by constructing their transfer matrices $S_0$. As a reminder, $S_0$ is the transfer matrix of the scattering medium that links its input and output surfaces, and differs from the scattering matrix $S$, which connects the input and output planes, encompassing the medium and its surrounding lenses.\\
For the thin medium case, $S_0$ is represented as a diagonal matrix of size $2601 \times 2601$. To generate this matrix, we first create a complex speckle pattern of size $51 \times 51$. This speckle pattern is a spatially random distribution of complex numbers with a controllable correlation length. The smaller the correlation length (noted $l_s$), the more complex the optical disorder. Figures~\ref{fig:SM_simu}e and f show the amplitude and phase component of such a complex speckle field. Next, the complex speckle field is reshaped into a $2601 \times 1$ vector and placed along the diagonal of the matrix $S_0$.\\
To model the transfer matrix $S_0$ of a thick scattering medium, we construct the matrix column by column.
Indeed, each column of $S_0$, reshaped as a $51 \times 51$ image, represents the complex field generated at the output surface of the medium in response to a point source located at the corresponding position on the input surface.
For a thin medium, it is assumed that a point at the input corresponds to a single point at the output, with only changes in amplitude and phase. In contrast, for a thick medium, the model accounts for coupling between an input point and multiple output points in a complex manner. This can be visualized as the `diagonal' of the matrix $S_0$ broadening to reflect the medium’s thickness.
Practically, $S_0$ is thus built column by column. For each column, a complex speckle field with a given correlation length $l_s$ is generated. This field is then multiplied element-wise by a Gaussian function of width $\sigma_s$, centered on the transverse input position corresponding to that column.
The thickness of the medium is controlled by varying the width $\sigma_s$ of the Gaussian. Repeating this procedure for all columns results in a matrix $S_0$ that models a thick scattering medium, characterized by the two parameters $l_s$ and $\sigma_s$ that define its complexity.
Figure~\ref{fig:SM_simu}g and h shows the amplitude and phase components of the speckle field associated with one column of the matrix $S_0$ modeling a thick medium.

\section{Experimental measurement of the transmission matrix $S_m$}
\label{matrixmeasure}

In our experiment, we experimentally measure the transmission matrix $S_m$, linking the SLM plane to the output plane, in the presence of the scattering medium in the system. For that, we illuminate the SLM using a collimated classical coherent light source (a SLED at $810$nm) and use the phase-shifting interferometric method described in Ref.~\cite{gigan_matrice_2010}. The active area of the SLM, comprising $400\times400$ (or $700\times700$ for the experiments in Figure 5) pixels, is subdivided into $32 \times32$ (or less) macro-pixels. On these macropixels, we sequentially display phase patterns corresponding to the Hadamard basis vectors. For each Hadamard pattern, we modulate its phase by applying different phase offsets $\theta_{SLM}\in [0,2\pi]$, resulting in a set of phase-shifted holographic measurements. The illumination around the active zone of the SLM is used as reference for the interferometric process. These allow us to extract both the amplitude and phase response of the medium in the output plane for each basis vector. After applying this process to all Hadamard basis elements, we perform a basis transformation to the SLM pixel basis in order to reconstruct the full complex matrix $S_m$. Its shape is \textit{number of pixels of the camera} $\times$ \textit{number of macro-pixels on the SLM}, which is typically $(128 \times 128) \times (32 \times 32)$.

\section{Experimental characterization of the two-photon `guide' state}
\label{charactState}

\begin{figure}
    \centering
    \includegraphics[width=0.5\linewidth]{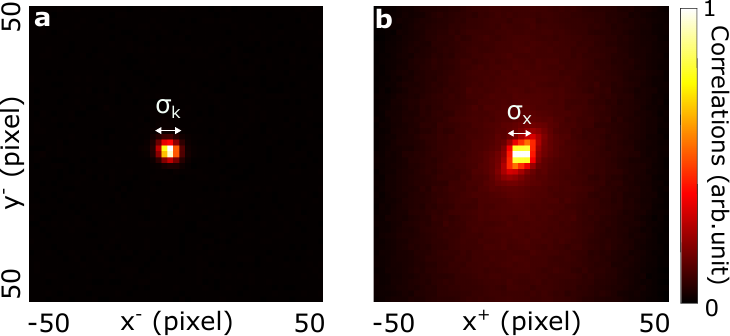}
    \caption{\textbf{Experimental characterization of the two-photon `guide' state.} Sum- and minus-coordinate projections of the second-order spatial correlation of photon pairs measured in the momentum basis \textbf{(a)} and in the position basis \textbf{(b)}, respectively. The peak widths in each image are proportional to $\sigma_{\vec{k}}$ and $\sigma_{\vec{r}}$, respectively. Their exact values, {$\sigma_{\vec{r}} \approx 13 \, \mu\text{m}$} and {$\sigma_{\vec{k}} \approx \num{4.7e3} \, \text{m}^{-1}$}, defined for the two-photon field in the crystal plane, are determined by applying a Gaussian fit to the images and accounting for the magnification factors between the crystal surface and the camera. An EMCCD camera was used to record these correlation images.
}
    \label{fig:FigX_SM_NF_FF}
\end{figure}

In this section, we detail the experimental characterization of the two-photon `guide' state used in the numerical optimization process (see Section~\ref{Optimpro}). By doing that, we also measure the position correlation width $\sigma_{\vec{r}}$, which depends solely on the crystal thickness, and is used in our simulations. Using Equation~\ref{eq3manu2} with a collimated Gaussian pump, the two-photon field in the crystal plane $\psi^{\text{c}}$ becomes:
\begin{equation}
    \psi^{\text{c}}({\vec{r_s},\vec{r_i}}) = \exp{ -\frac{\sigma_{\vec{k}}^2 |\vec{r_i}+\vec{r_s}|^2}{4}} \exp{\frac{-|\vec{r_i}-\vec{r_s}|^2 }{4\sigma_{\vec{r}}^2}},
\label{GaussianModel21}
\end{equation}
where $\vec{r_i}$ and $\vec{r_s}$ are the transverse spatial coordinate of the idler and signal photons in the crystal plane, $\sigma_{\vec{r}}$ and $\sigma_{\vec{k}}$ are the position and momentum correlation widths, respectively. We can also express the two-photon field in the momentum basis by performing a Fourier transform: 
\begin{equation}
    \tilde{\psi}^{\text{c}}({\vec{k_s},\vec{k_i}}) = \exp{ \frac{ - |\vec{k_i}+\vec{k_s}|^2}{4 \sigma_{\vec{k}}^2}} \exp{\frac{-\sigma_{\vec{r}}^2 |\vec{k_i}-\vec{k_s}|^2} {4}},
\label{GaussianModel23}
\end{equation}
where $\vec{k_i}$ and $\vec{k_s}$ are the transverse spatial momentum of the idler and signal photons in the crystal plane. Following the methodology of Refs.~\cite{Moreau2012,Edgar2012}, we experimentally measure both position and momentum correlation widths using an EMCCD camera.\\ 
\\
\noindent \textbf{Momentum correlation width $\sigma_{\vec{k}}$:} To evaluate the momentum correlation width, the crystal is Fourier imaged onto the EMCCD camera. {To do that, the optical configuration used is the same as that described in Section~\ref{setupconf}, with $f_5 = 125\,\text{mm}$, but $f_c$ is replaced by two lenses in a confocal configuration, $150$mm–$50$mm, resulting in an effective focal length from the crystal to the EMCCD camera of $26\,\text{mm}$}. 
The second-order spatial correlation function is then measured and projected onto the sum-coordinate axis, as shown in Figure~\ref{fig:FigX_SM_NF_FF}a. 
The width of the correlation peak is proportional to the momentum correlation width $\sigma_{\vec{k}}$. 
Using a Gaussian fit and taking into account the effective focal length, we find {$\sigma_{\vec{k}} \approx \num{4.7e3} \, \text{m}^{-1}$}.  \\
\\
\noindent \textbf{Position correlation width $\sigma_{\vec{r}}$}: To measure the position correlation width, we image the crystal surface onto an EMCCD camera. {To do that, we use a configuration similar to the one described in Section~\ref{setupconf}, with $f_5 = 125\,\text{mm}$, but the imaging system $f_c$ consists of a first lens ($150\,\text{mm}$) in a 2$f$-configuration (Fourier transform), followed by a $25\,\text{mm}$ lens in a single-lens imaging configuration. This setup provides a total magnification of 1.9 from the crystal to the camera.}
 The second-order spatial correlation function is then measured and projected onto the minus-coordinate axis, as shown in Figure~\ref{fig:FigX_SM_NF_FF}b. The width of the correlation peak is proportional to the position correlation width $\sigma_{\vec{r}}$. Using a Gaussian fit and taking into account the magnification factor, we find {$\sigma_{\vec{r}} \approx 13 \, \mu$m}.  

\section{Additional details on the numerical optimization process}
\label{Optimpro}

As explained in the manuscript, it is faster in our case to perform the optimization numerically to determine the tailored scattering matrix $S'$ (although it can also be done in principle experimentally, as demonstrated in Section~\ref{expopt}). As detailed in Methods, this numerical optimization process use an experimentally measured transmission matrix $S_m$ (Section~\ref{matrixmeasure}) and an experimentally characterized two-photon `guide' state in the SLM plane (Section~\ref{charactState}). \\
\\
\noindent \textbf{Discretization of $S_m$ and $\Psi^{\text{SLM}}$:} The transmission matrix $S_m$, measured as described in Section~\ref{matrixmeasure} using the EMCCD camera, has dimensions $16384 \times 1024$. Since this matrix is too large for efficient optimization, we use a reduced portion of it with size $2500 \times 1024$. Note that the matrix $S_m$ encompasses both the scattering medium and the optical elements following the SLM, effectively accounting for scattering and Fourier imaging onto the camera.\\
The field of the two-photon `guide' state in the SLM plane is modeled using Equation~\ref{GaussianModel21}, taking into account the magnification $M''$ between the crystal plane and the SLM plane:
\begin{equation}
    \psi^{\text{SLM}}({\vec{r_i},\vec{r_s}}) = \exp{ -\left(\frac{\sigma_{\vec{k}}}{M''} \right)^2 |\vec{r_i}+\vec{r_s}|^2} \exp{\frac{-|\vec{r_i}-\vec{r_s}|^2 }{(M''\sigma_{\vec{r}})^2}},
\label{GaussianModel24}
\end{equation}
where $M'' = 4.3$ is the magnification between the crystal plane and the SLM plane, $\sigma_{\vec{r}} = 13\, \mu\text{m}$, and $\sigma_{\vec{k}} = 4.7 \times 10^{3} \, \text{m}^{-1}$. The correlation width in the SLM plane is therefore approximately $M'' \sigma_{\vec{r}} \approx 62\, \mu\text{m}$. Knowing the actual size of one SLM macropixel ($100\, \mu\text{m}$ for the results in Figure 3 and $175\, \mu\text{m}$ for those in Figure 5), we discretize the function $\psi^{\text{SLM}}$ given in Equation~\ref{GaussianModel24} over a $32 \times 32$ spatial grid to obtain the two-photon matrix $\Psi^{\text{SLM}}$. \\
\\
\noindent \textbf{Numerical optimization process:} {As explained in Methods, the matrix $S'$ can de decomposed as follow: 
\begin{equation}
    S'=S_m D_{SLM} F_{f_1},
\end{equation}
where $D_{SLM}$ is a diagonal matrix modeling the action of the SLM, composed of $32 \times 32 = 1024$ pure phase terms on its diagonal. In practice, to minimize numerical errors introduced by the discrete Fourier transform $F_{f_1}$, we simulate the propagation directly of the experimentally measured state in the SLM plane, $\Psi^{\text{SLM}}$, using the following formula: 
\begin{equation}
    \Psi^{\text{out}} = S_m D_{\text{SLM}} \Psi^{\text{SLM}} (S_m D_{\text{SLM}}) ^t.
\end{equation}
The optimization target is the correlation value at the central pixel of the sum-coordinate projection of $\Psi^{\text{out}}$, noted $\Gamma^+_0$. To increase it, we iteratively modify the SLM phase using a random partitioning algorithm~\cite{vellekoop_phase_2008}.} Each iteration proceeds as follows:
\begin{enumerate}
\item A random subset of half the SLM pixels is selected. These pixels are sequentially assigned six phase values $\theta_{\text{SLM}} \in \{0, \pi/3, 2\pi/3, \pi, 4\pi/6, 5\pi/3 \}$.
\item For each phase value, the quantum state is propagated through the corresponding SLM pattern and scattering medium using:
\begin{equation}
\Psi^{\text{out}} = S_m D_{\text{SLM}} \Psi^{\text{SLM}} (S_m D_{\text{SLM}}) ^t.
\end{equation}
\item The sum-coordinate projection of $\Psi^{\text{out}}$ is calculated and the correlation value at the center pixel $\Gamma^+_0$ is recorded. 
\item The resulting six values of $\Gamma^+_0$ obtained for the six different phases are fit to a double cosine function of the form, (see details in Ref. ~\cite{courme2025nonclassicaloptimizationcomplexmedia}):
\begin{equation}
\Gamma^+_0(\theta) = a\cos(\theta + \theta_a) + b\cos(2\theta + \theta_b) + c,
\label{double_cosine}
\end{equation}
where $a$, $b$, $c$, $\theta_a$, and $\theta_b$ are fitting parameters, and $\theta$ is the applied phase shift.
\item The phase shift $\theta_{opt}$ that maximizes this function is then applied to all pixels in the selected group, and the process is repeated for the next random subset.
\end{enumerate}

{At each iteration of the optimization, the value of the central peak in the correlation image increases progressively. As shown in Figure 4g of the manuscript, the process typically requires around 4000 steps to reach a saturation point or plateau using $32 \times 32$ SLM modes.
Once the optimization converges, it yields a phase pattern $D_{\text{SLM}}$ that maximizes the correlation peak. This optimized phase pattern effectively implements the transformation $S'$. It is then programmed onto the SLM for use in the real experiment.
Note that this approach assumes that experimental conditions - such as lenses alignment and the scattering medium - remain sufficiently stable throughout the entire duration of the numerical optimization, which takes several tens of minutes.}

\section{Preliminary results on experimental optimization}
\label{expopt} 

\begin{figure}[t!]
    \centering
    \includegraphics[width=0.8\linewidth]{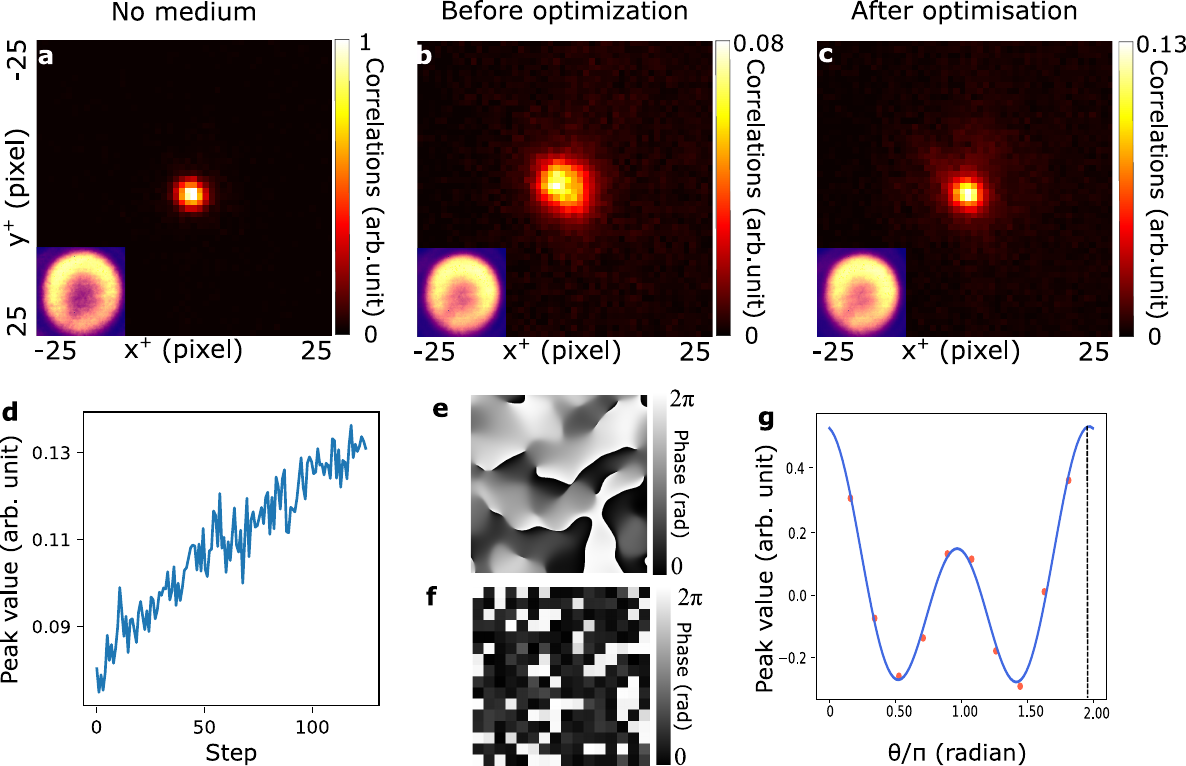}
    \caption{\textbf{Preliminary results of experimental optimization}. Correlation images measured with the Tpx3cam with an acquisition time of $3$ seconds without disordered medium \textbf{(a)}, in presence of the disordered medium \textbf{(b)} and after the optimization \textbf{(c)}. \textbf{d}, Optimization curve showing the central peak value $\Gamma^+_0$ as a function of the number of steps of the optimization, which was stopped after $130$ steps ($17$ hours). \textbf{e}, Phase disorder displayed by the SLM. \textbf{f}, Last phase mask obtained. \textbf{g}, Target pixel value $\Gamma^+_0$ oscillating with the phase-shifting $\theta$ value. Red spots correspond to the experimental value and the blue curve corresponds to the fit according to Equation~\ref{double_cosine}. The vertical dashed line points to the optimal phase value retained for this optimization step of the partitioning algorithm.}
    \label{fig:SM_opti}
\end{figure}

As described in Section~\ref{Optimpro}, the optimization process to find $S'$ is, for practical reasons, carried out computationally. However, following the experimental approach presented in Ref.~\cite{courme2025nonclassicaloptimizationcomplexmedia}, this optimization can also be implemented experimentally. In the following, we present preliminary results of such an experimental optimization using the Tpx3Cam camera.

In this proof-of-principle experiment, the SLM is initialized with a random phase pattern, thereby serving both as the disordered medium and as the control tool. A collimated pump beam is directed onto the crystal, and the resulting correlation image - specifically, its central value $\Gamma^+_0$ - is measured using the Tpx3Cam camera. Figure~\ref{fig:SM_opti}a shows the reference correlation image obtained without any SLM phase disorder, while Figure~\ref{fig:SM_opti}b displays the broadened correlation peak observed in the presence of the optical disorder. The random SLM phase mask used to generate the disorder is shown in Figure~\ref{fig:SM_opti}e.

The experimental optimization strategy follows the same method as in the numerical case presented in Section~\ref{Optimpro}: half of the SLM pixels are selected and sequentially phase-shifted across $10$ equally spaced values between $0$ and $2\pi$ (instead of $6$ in the numerical case). For each phase value, a correlation image is acquired, and the central correlation value $\Gamma^+_0$ is used as feedback. These values are then fitted using the double cosine model described in Equation~\ref{double_cosine}. The modulation of $\Gamma^+_0$, along with the fitted curve, is shown in Figure~\ref{fig:SM_opti}g. The phase corresponding to the maximum of the fit (indicated by a vertical line) is retained, and the process is repeated with another randomly chosen group of pixels.

The optimization is performed directly on the Tpx3Cam camera, using a $16 \times 16$ SLM phase mask. Each correlation image is acquired over a 3-second exposure. With 10 phase steps per group, each optimization point takes approximately 450 seconds. The optimization was stopped after 130 steps, leading to a total runtime of about 17 hours. In comparison, the numerical optimization for a $16 \times 16$ mask takes around 30 minutes, plus 15 minutes for the transmission matrix measurement. After optimization, the tailored disorder produces the sharpened correlation image shown in Figure~\ref{fig:SM_opti}c. Figure~\ref{fig:SM_opti}d shows the experimental optimization curve, tracking the central correlation pixel value as a function of iteration number. Figure~\ref{fig:SM_opti}f displays the final phase pattern obtained at the last step.
Note that the optimization was deliberately stopped before reaching a plateau to avoid prolonged exposure of the Tpx3Cam camera’s photon intensifier. Nonetheless, these preliminary results are sufficient to demonstrate the feasibility of performing the optimization entirely using the camera, without relying on numerical simulations or classical light.

\section{Approximation $\sigma_{\vec{r}} \rightarrow 0$ not satisfied in our experiment} 
{The approximation $\sigma_{\vec{r}} \rightarrow 0$, also known as the `thin crystal' approximation~\cite{abouraddy_entangled-photon_2002}, is not satisfied in our work. This is demonstrated by the experimental results shown in Figure~\ref{fig:SM_opti}g. Indeed, the curve in Figure~\ref{fig:SM_opti}g shows the modulation of $\Gamma^+_0$, the central value of the output correlation image, when a phase shift $\theta$ is applied to half of the SLM pixels, randomly selected. As detailed in Section~\ref{Optimpro} and in Ref.~\cite{courme2025nonclassicaloptimizationcomplexmedia}, this modulation follows the form:
\begin{equation}
 \Gamma^+_0(\theta) = a\cos(\theta + \theta_a) + b\cos(2\theta + \theta_b) + c.   
\end{equation}
This behavior results from the propagation of the two-photon state from the SLM to the output plane, described by:
\begin{equation}
\Psi^{\text{out}} = S_m D_{\text{SLM}} \Psi^{\text{SLM}} (S_m D_{\text{SLM}})^t.
\end{equation}

If we assume $\sigma_{\vec{r}} \rightarrow 0$, however, then the two-photon state in the SLM plane (Equation~\ref{GaussianModel24}) simplifies to:
\begin{equation}
\psi^{\text{SLM}}(\vec{r_i}, \vec{r_s}) \approx \exp\left[ -\left( \frac{\sigma_{\vec{k}}}{M''} \right)^2 |\vec{r_i} + \vec{r_s}|^2 \right] \delta(\vec{r_i} - \vec{r_s}),
\end{equation}
which implies that the two-photon matrix $\Psi^{\text{SLM}}$ can be approximated by a diagonal matrix. If that were the case, the modulation of $\Gamma^+_0(\theta)$ would take the form $b\cos(2\theta + \theta_b) + c$, excluding the $\cos(\theta)$ term. However, the experimental measurement shown in Figure~\ref{fig:SM_opti}g clearly displays a contribution from the $\cos(\theta)$ term, indicating that the approximation $\sigma_{\vec{r}} \rightarrow 0$ does not hold in our experiment. 

This demonstrates that, in order to determine an experimentally valid transformation $S'$, one cannot rely on this approximation and the theoretical framework built upon it. This limitation justifies the use of the optimization-based method employed in our experiment. Nevertheless, analyzing this limiting case, as done in the main text, provides useful guidance in the search for such solutions.
}

\section{Comparison between the experimental solutions $S'$ and the ones derived from Equation (5) of the manuscript}
\label{comparisonsolution}

\begin{figure}[t!]
    \centering
    \includegraphics[width=0.95\linewidth]{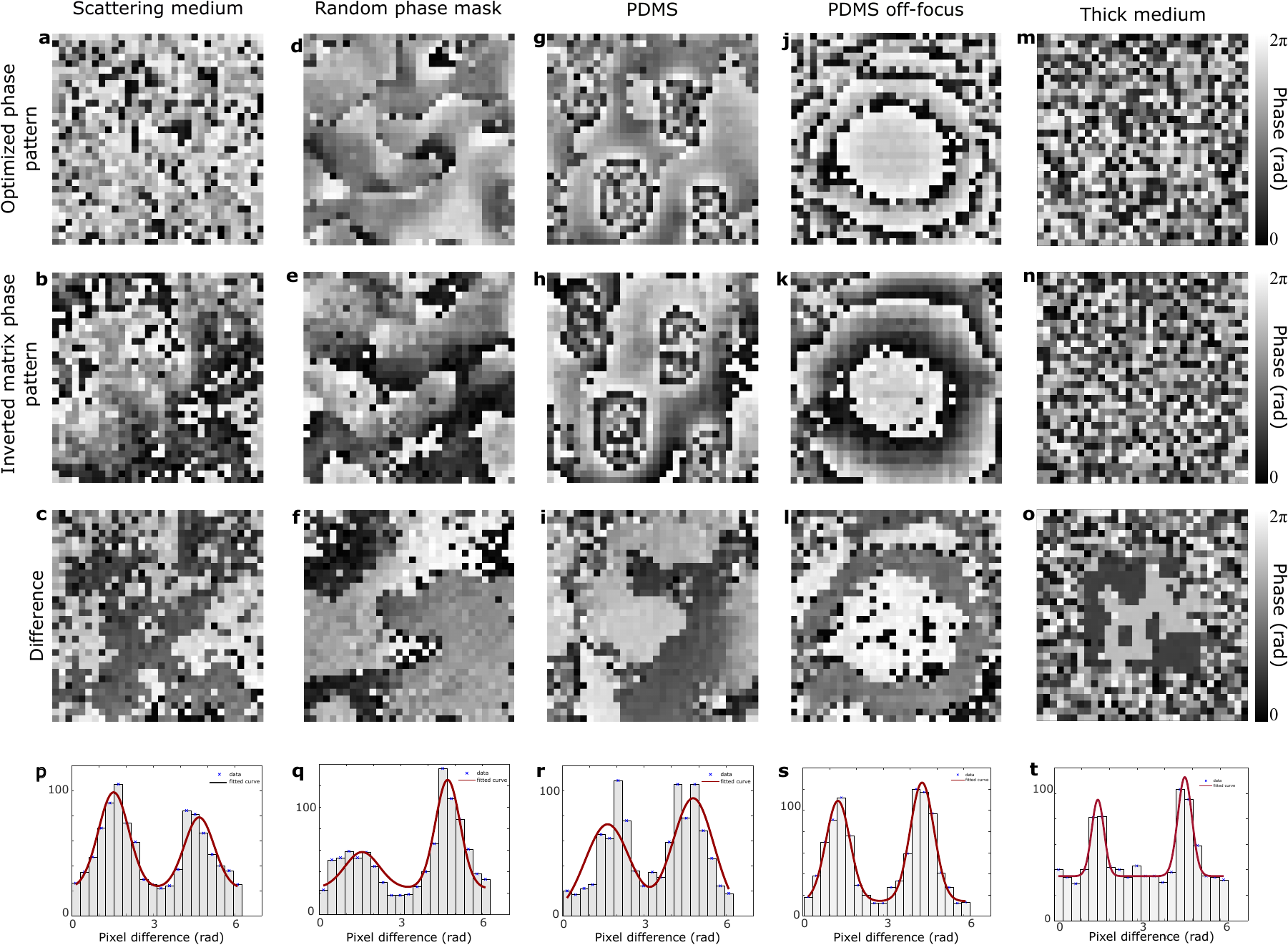}
    \caption{\textbf{Optimal phase patterns obtained after optimization}. Optimized phase pattern obtained for the parafilm layer \textbf{(a)}, the random phase mask \textbf{(d)}, the PDMS \textbf{(g)}, the PDMS off-focus \textbf{(j)} and the simulated thick scattering medium \textbf{(m)}. These patterns implement the transformation $S'$. Phase patterns obtained by inverting the scattering matrix of the parafilm layer \textbf{(b)}, the random phase mask \textbf{(e)}, the PDMS \textbf{(h)}, the PDMS off-focus \textbf{(k)} and the simulated thick scattering medium \textbf{(n)}. These patterns implement the identity transformation.
    Difference between the optimized and the inverted patterns for the parafilm layer \textbf{(c)}, the random phase mask \textbf{(f)}, the PDMS \textbf{(i)}, the PDMS off-focus \textbf{(l)} and the simulated thick scattering medium \textbf{(o)}. 
    All phase pattern are made of $32\cross 32$ SLM macropixels. 
    \textbf{p-t}, Pixel-wise histogram of the difference phase pattern shown above in \textbf{c-o} respectively. The distribution are fitted by a double Gaussian centered on two values separated by $\pi$ (red).}
    \label{fig:SM_slmpatterns}
\end{figure}

{To compare the experimentally optimized solutions $S'$ with the theoretical predictions from Equation (5) of the manuscript, we analyze the corresponding phase patterns on the SLM obtained at the end of the optimizations.}
Figure~\ref{fig:SM_slmpatterns} presents, in the top row, the optimal SLM phase patterns - implementing $S'$ - obtained for the five complex media discussed in the main text: from left to right, the scattering medium, the random phase disorder, the PDMS sample, the PDMS out of focus and the thick medium (simulation). 
The second row of Figure~\ref{fig:SM_slmpatterns} displays the corresponding phase patterns computed by inverting the transmission matrix~\cite{gigan_matrice_2010}, which effectively implement a transformation close to the the identity transformation. 
The third row shows the difference between the two patterns for each considered medium. 

If these solutions, found by optimization, corresponded to the theoretical solutions predicted by Equation (5) of the manuscript, then the difference SLM phase patterns shown in the third row of Figure~\ref{fig:SM_slmpatterns} would match the phase of a function of the form $\text{sign}(g)$, where $g$ is an arbitrary function. In other words, these patterns would consist exclusively of randomly distributed phases of either $0$ or $\pi$.
However, the phase patterns shown in the third row clearly demonstrate that this is not the case experimentally. This discrepancy arises from the fact that the assumption $\sigma_{\vec{r}} \rightarrow 0$ is not satisfied under our experimental conditions.

{Nevertheless, the patterns exhibit shared structural features. In particular, the bottom row of Figure~\ref{fig:SM_slmpatterns} displays the distribution of pixel-wise phase differences between the two patterns obtained for each medium. In all cases, two distinct peaks separated by $\pi$ are clearly visible. The phases are therefore, on average, separated by a phase difference of $\pi$.
This shows that, although the optimization cannot drive the system exactly toward the ideal solutions composed of $0$ and $\pi$ phases predicted by Equation (5), it nonetheless converges toward solutions that closely resemble them.}

\section{Comparison with classically-correlated photon pairs} 

In this section, we demonstrate that entanglement is essential to our approach, and that classical correlations alone are not sufficient. In particular, classical correlations fail at two critical stages: (i) implementing the optimization procedure required to determine the transformation $S'$, and (ii) transmitting the image, even when $S'$ is assumed to be known. To highlight these limitations, we perform simulations using two-photon mixed separable states, which are constructed to reproduce exactly the same correlation image as the entangled photon pairs in the input plane. Despite this, the separable state fails to enable either (i) or (ii), underscoring the crucial role of quantum entanglement in our protocol.\\
\\
\noindent \textbf{(i) Insufficiency to drive the optimization toward a non-trivial solution $S'$:} To find $S'$ in our experiment, we use an optimization process process that employs a `guide' state at the input whose the associated two-photon wavefunction is modeled by a double Gaussian:
\begin{equation}
    {\psi}^{\text{in}}({\vec{r_i},\vec{r_s}}) = \exp{ \frac{ - \pi^2 |\vec{r_i}+\vec{r_s}|^2}{4 \lambda_p^2 f_1^2 \sigma_{\vec{k}}^2}} \exp{\frac{-\pi^2 \sigma_{\vec{r}}^2 |\vec{r_i}-\vec{r_s}|^2} {4 \lambda_p^2 f_1^2}}.
\end{equation}
Theoretically, one can construct a mixed state, denoted $\rho_0$, that is separable but reproduces exactly the same spatial correlations as $\psi^{\text{in}}$ in the input plane. This state is defined as follows:
\begin{equation}
    \rho_0 = \int p_{\vec{R}} \ket{\psi_0^{\vec{R}}}\bra{\psi_0^{\vec{R}}} d \vec{R},
\end{equation}
where $p_{\vec{R}} = A$ are constant probabilities and $\ket{\psi_0^{\vec{R}}}$ is a pure two-photon state defined by a separable wavefuntion $\psi_0^{\vec{R}}(\vec{r_i},\vec{r_s}) =  \phi_0^{\vec{R}}(\vec{r_i}) \chi_0^{\vec{R}}(\vec{r_s}) $. These functions are written as:
\begin{eqnarray}
    \phi_0^{\vec{R}}(\vec{r_i}) &=&  \exp{ \frac{ - \pi^2 |\vec{r_i}+\vec{R}|^2}{4 \lambda_p^2 f_1^2 \sigma_{\vec{k}}^2}} \exp{\frac{-|\vec{r_i}-\vec{R}|^2 \pi^2 \sigma_{\vec{r}}^2 }{4 \lambda_p^2 f_1^2}} \\
    \chi_0^{\vec{R}}(\vec{r_s}) &=& \delta \left( \vec{r_s} - \vec{R} \right).
\end{eqnarray}
With this definition of $\rho_0$, we show below that it yields exactly the same correlation image as the entangled state used in our experiment:
\begin{eqnarray}
\Gamma^+(\vec{r_+})  &=& \int G^{(2)}(\vec{r_+}-\vec{r},\vec{r}) d \vec{r} \\
&=& \iint p_{\vec{R}} |\psi_0^{\vec{R}}(\vec{r_+}-\vec{r},\vec{r})|^2  d \vec{R} d \vec{r} \\
&=& A \iint \exp{ \frac{ - \pi^2 |\vec{r_+}-\vec{r}+\vec{R}|^2}{2 \lambda_p^2 f_1^2 \sigma_{\vec{k}}^2}} \exp{\frac{-|\vec{r_+}-\vec{r}-\vec{R}|^2 \pi^2 \sigma_{\vec{r}}^2 }{2 \lambda_p^2 f_1^2}} \delta \left( \vec{r} - \vec{R} \right)  d \vec{R} d \vec{r}  \\
&=& A \int  \exp{ \frac{ - \pi^2 |\vec{r_+}|^2}{2 \lambda_p^2 f_1^2 \sigma_{\vec{k}}^2}} \exp{\frac{-|\vec{r_+}-2\vec{r}|^2 \pi^2 \sigma_{\vec{r}}^2 }{2 \lambda_p^2 f_1^2}}   d \vec{r}  \\
&=& B \exp{ \frac{ - \pi^2 |\vec{r_+}|^2}{2 \lambda_p^2 f_1^2 \sigma_{\vec{k}}^2}},
\end{eqnarray} 
where $B$ is a normalization constant. 
\begin{figure}[t!]
    \centering    \includegraphics[width=0.6\linewidth]{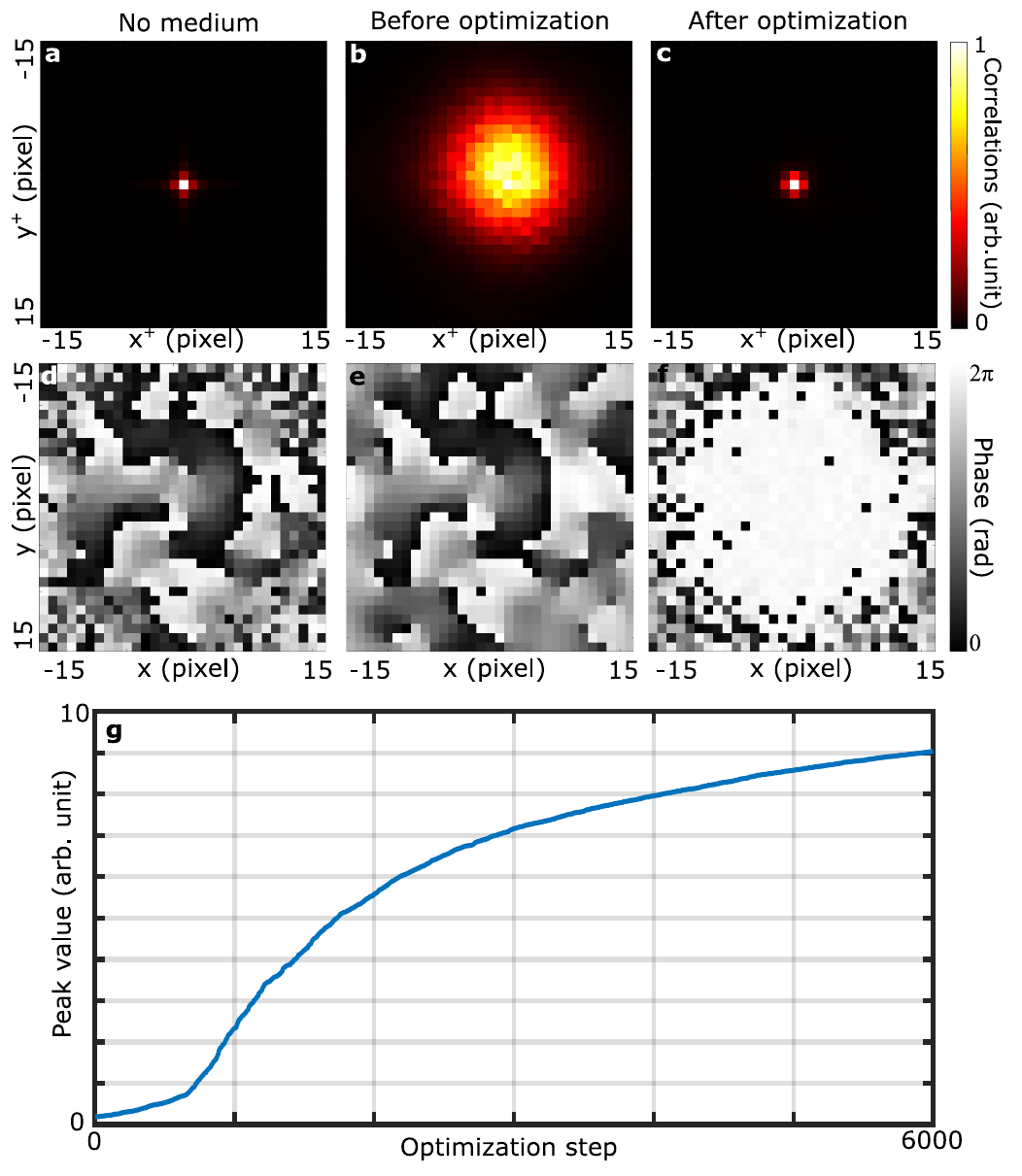}
    \caption{\textbf{Optimization with a classically-correlated `guide' state.} Simulated correlation image of the separable state $\rho$ without medium \textbf{(a)}, after propagation through a thin scattering medium before the optimization \textbf{(b)} and after the optimization \textbf{(c)}. \textbf{d,} Optimal phase mask obtained at the end of the optimization. \textbf{e,} Phase mask calculated from the scattering matrix that approximates the identity transformation. \textbf{e,} Difference between the masks. \textbf{f,} Evolution of the target correlation value $\Gamma^+_0$ as a function of the number of optimization steps.}
    \label{fig:SM_corrClassique1}
\end{figure}

\begin{figure}[t!]
    \centering
    \includegraphics[width=1\linewidth]{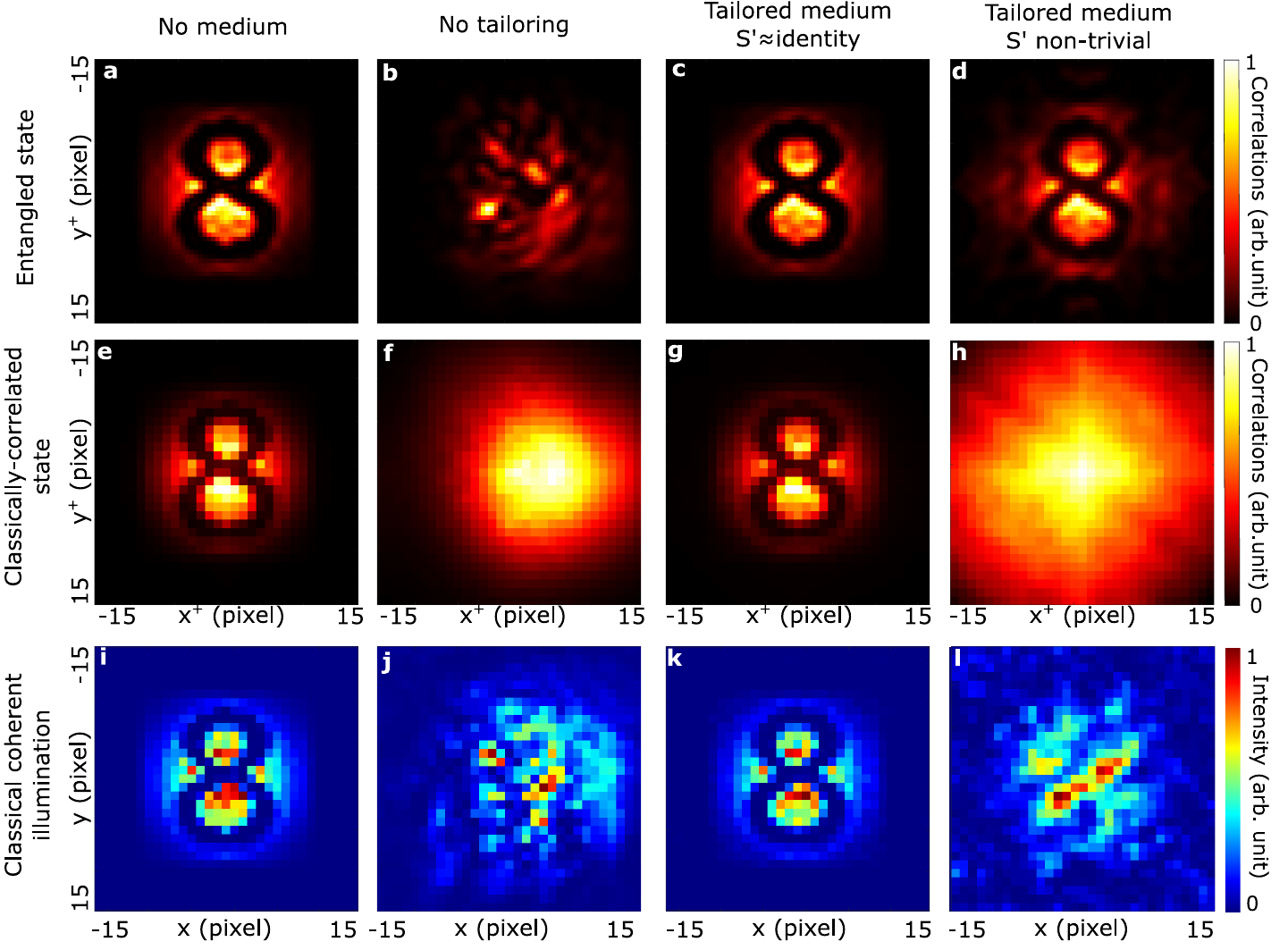}
    \caption{\textbf{Propagation of an object encoded in classical correlations.} Simulated correlation images of a binary digit object ‘8’ encoded in the entangled state at the input (i.e. without a medium) \textbf{(a)}, after propagation through a thin scattering medium \textbf{(b)}, after propagation through a transformation $S'$ that approximates the identity \textbf{(c)}, and after propagation through a non-trivial transformation \textbf{(d)}. Simulated correlation images of a binary digit object ‘8’ encoded in the classically-correlated state $\rho_t$ at the input (i.e. without a medium) \textbf{(e)}, after propagation through a thin scattering medium \textbf{(f)}, after propagation through a transformation $S'$ that approximates the identity \textbf{(g)}, and after propagation through a non-trivial transformation \textbf{(h)}. Simulated intensity images of a binary digit object ‘8’ under classical illumination at the input (i.e. without a medium) \textbf{(i)}, after propagation through a thin scattering medium \textbf{(j)}, after propagation through a transformation $S'$ that approximates the identity \textbf{(k)}, and after propagation through a non-trivial transformation \textbf{(l)}.
}
    \label{fig:SM_corrClassique3}
\end{figure}

To study the behavior of this separable state, we perform numerical simulations.
Figure~\ref{fig:SM_corrClassique1}a shows the correlation image of the classically correlated state $\rho_0$ in the input plane (i.e. without a scattering medium), which exhibits a sharp peak at its center, just like the entangled state.
When propagating through a thin scattering medium, a diffuse halo appears in the correlation image, as shown in Figure~\ref{fig:SM_corrClassique1}b. 
Using the optimization approach, we iterate on the SLM to increase the correlation value at the central peak $\Gamma^+_0$, thereby recovering a sharp peak in the correlation image at the end of the process, as shown in Figure~\ref{fig:SM_corrClassique1}c. 
The evolution of $\Gamma^+_0$ as a function of the optimization step is shown in Figure~\ref{fig:SM_corrClassique1}g. 
As expected, the separable state $\rho_0$ indeed allows for implementing an optimization.

However, we observe that the solution thus obtained, $S'$, is in fact the identity and not a non-trivial transformation. To demonstrate this, one can, for instance, compare the SLM pattern obtained at the end of the optimization (Fig.~\ref{fig:SM_corrClassique1}d) with the one computed directly from the transmission matrix to implement the identity (Fig.~\ref{fig:SM_corrClassique1}e) (see Section~\ref{comparisonsolution} for details about this SLM phase pattern comparison). 
The subtraction of these two SLM patterns is shown in Figure~\ref{fig:SM_corrClassique1}f, clearly indicating that they are identical (except for the edges, which appear noisy and correspond to regions of the SLM receiving lower intensity). 

These results show that, when the optimization process to find $S'$ is performed using such a mixed state $\rho_0$ as the `guide' state, the system systematically converges toward the identity and does not allow to find of non-trivial solutions.
We note that this result is not an isolated case: numerous simulations were carried out to confirm this conclusion. Moreover, this point has already been discussed in Ref.~\cite{courme2025nonclassicaloptimizationcomplexmedia}. \\
\\
\noindent \textbf{(ii) Insufficiency to transmit the image if a non-trivial solution $S'$ is implemented.}
{Let us now assume that a non-trivial solution $S'$ has already been found. In this case, it is also possible to prepare a mixed separable state, denoted $\rho_t$, which reproduces the same correlation image - encoding the object - as that produced by entangled photon pairs in the input plane of our experiment. To do so, $\rho_t$ must be constructed as follows:
\begin{equation}
    \rho_t = \int p_{\vec{R}} \ket{\psi^{\vec{R}}}\bra{\psi^{\vec{R}}} d \vec{R},
\end{equation}
where $p_{\vec{R}} = A$ are constant probabilities and $\ket{\psi^{\vec{R}}}$ is a pure two-photon state defined by a separable wavefuntion $\psi^{\vec{R}}(\vec{r_i},\vec{r_s}) =  \phi^{\vec{R}}(\vec{r_i}) \chi^{\vec{R}}(\vec{r_s}) $. These functions are written as:
\begin{eqnarray}
    \phi^{\vec{R}}(\vec{r_i}) &=&  {t} \left(\frac{\vec{r_i}+\vec{R}}{M} \right) \exp{\frac{-|\vec{r_i}-\vec{R}|^2 \pi^2 \sigma_{\vec{r}}^2 }{4 \lambda_p^2 f_1^2}} \\
    \chi^{\vec{R}}(\vec{r_s}) &=& \delta \left( \vec{r_s} - \vec{R} \right).
\end{eqnarray}
With this definition of $\rho_t$, we show below that it yields exactly the same object-encoded correlation image as the entangled state used in our experiment:
\begin{eqnarray}
\Gamma^+(\vec{r_+})  &=& \int G^{(2)}(\vec{r_+}-\vec{r},\vec{r}) d \vec{r} \\
&=& \iint p_{\vec{R}} |\psi^{\vec{R}}(\vec{r_+}-\vec{r},\vec{r})|^2  d \vec{R} d \vec{r} \\
&=& A \iint \left|{t} \left(\frac{\vec{r_+}-\vec{r}+\vec{R}}{M} \right) \right|^2 \exp{\frac{-|\vec{r_+}-\vec{r}-\vec{R}|^2 \pi^2 \sigma_{\vec{r}}^2 }{2 \lambda_p^2 f_1^2}} \delta \left( \vec{r} - \vec{R} \right)  d \vec{R} d \vec{r}  \\
&=& A \int  \left|{t} \left(\frac{\vec{r_+}}{M} \right)\right|^2 \exp{\frac{-|\vec{r_+}-2\vec{r}|^2 \pi^2 \sigma_{\vec{r}}^2 }{2 \lambda_p^2 f_1^2}}   d \vec{r}  \\
&=& B \left|{t} \left(\frac{\vec{r_+}}{M} \right)\right|^2,
\end{eqnarray} 
where $B$ is a normalization constant.}

To study the behavior of this separable state, we perform numerical simulations.
First, Figure~\ref{fig:SM_corrClassique1}e shows the correlation image of the classically correlated state $\rho_t$ in the input plane (i.e. without a scattering medium), which reveals the object, just like the entangled state (Fig.~\ref{fig:SM_corrClassique1}a). The intensity image of the object obtained under classical illumination is also shown in Figure~\ref{fig:SM_corrClassique1}i.
After propagating through a thin scattering medium, all images become diffused: a diffuse halo appears in the correlation image of $\rho_t$ (Fig.~\ref{fig:SM_corrClassique1}f), whereas speckle patterns appear for the entangled state (Fig.~\ref{fig:SM_corrClassique1}b) and the classical illumination (Fig.~\ref{fig:SM_corrClassique1}j).
We then program the transformation $S'$. 
First, we consider a trivial case where the matrix $S'$ is close to the identity, which is in fact the solution obtained with the state $\rho\_0$ in the previous subsection. As expected in this case, the object reappears at the output of the medium regardless of the input state (Figs.~\ref{fig:SM_corrClassique1}c, g, and k).
Next, we consider the case of interest, where $S'$ is a non-trivial solution obtained using an entangled `guide' state. In this case, as expected, the image encoded in the entangled state reappears at the output (Fig.~\ref{fig:SM_corrClassique1}d), whereas this is not the case for the image of the object illuminated with classical light (Fig.~\ref{fig:SM_corrClassique1}i). More importantly, as shown in Figure~\ref{fig:SM_corrClassique1}f, the image of the object encoded in the separable state $\rho_t$ does not reappear at the output.

This demonstrates that classical correlations alone are not sufficient to preserve the object through the non-trivial solution $S'$, and that quantum entanglement is indeed necessary. Note that this observation is not an isolated case; numerous simulations have been performed to confirm this result.

\section{Details on the number of SLM modes controlled during optimization and the resulting image quality at $S'$ output.}

\begin{figure}[t!]
    \centering
    \includegraphics[width=0.65\linewidth]{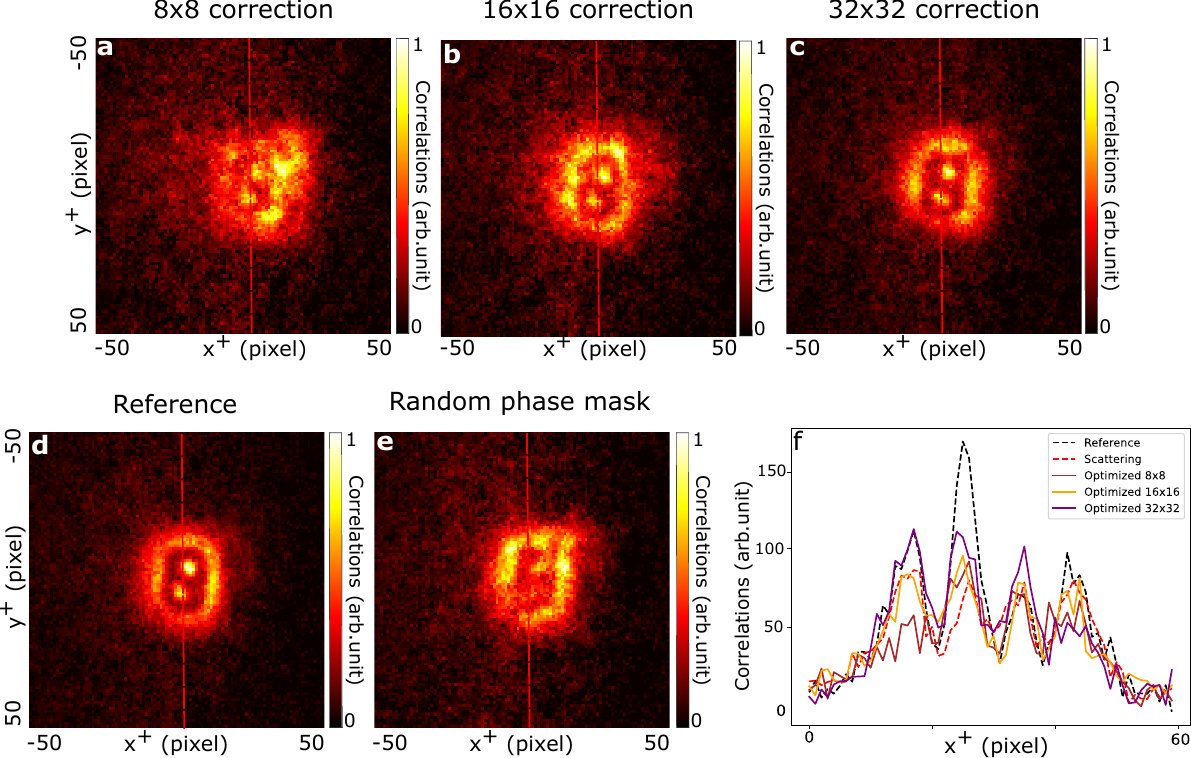}
    \caption{\textbf{Effect of the number of SLM macropixels on output image quality {after tailoring the optical disorder}}. \textbf{a–c}, Experimental correlation images obtained at the output of a tailored optical system after applying optimized phase patterns with $8\times8$, $16\times16$, and $32\times32$ SLM macropixels, respectively. \textbf{d}, Reference correlation image without disorder. \textbf{e}, Correlation image through the disorder without correction. \textbf{f}, Vertical cut (corresponding to the vertical red line) showing that both contrast and resolution increase with the number of SLM macropixels used for tailoring the system. All correlation images correspond correspond to an acquisition time of $4$ seconds.}
    \label{fig:disorderPixelsize}
\end{figure}

{Similar to classical wavefront shaping~\cite{katz_looking_2012,vellekoop_focusing_2007}, increasing the number of macropixels controlled by the SLM to find the non-trivial transformation $S'$ improves the quality of the correlation image obtained at the output.}  Figure~\ref{fig:disorderPixelsize} shows experimental correlation images at the output of optical systems tailored with optimized phase patterns using different macropixel sizes: $8\times8$ (a), $16\times16$ (b), and $32\times32$ (c). As the number of SLM macropixels increases, the correction becomes more precise. For comparison, Figure~\ref{fig:disorderPixelsize}d shows the reference correlation image without disorder, while Figure~\ref{fig:disorderPixelsize}e shows the correlation image through the disorder without correction. Figure~\ref{fig:disorderPixelsize}f displays a vertical cut (indicated by a red line on each correlation image), highlighting the improvement in contrast and resolution with increasing numbers of SLM pixels used for correction.

{\section{Details on the number of existing non-trivial solutions $S'$}}

\begin{figure}[t!]
    \centering
\includegraphics[width=\linewidth]{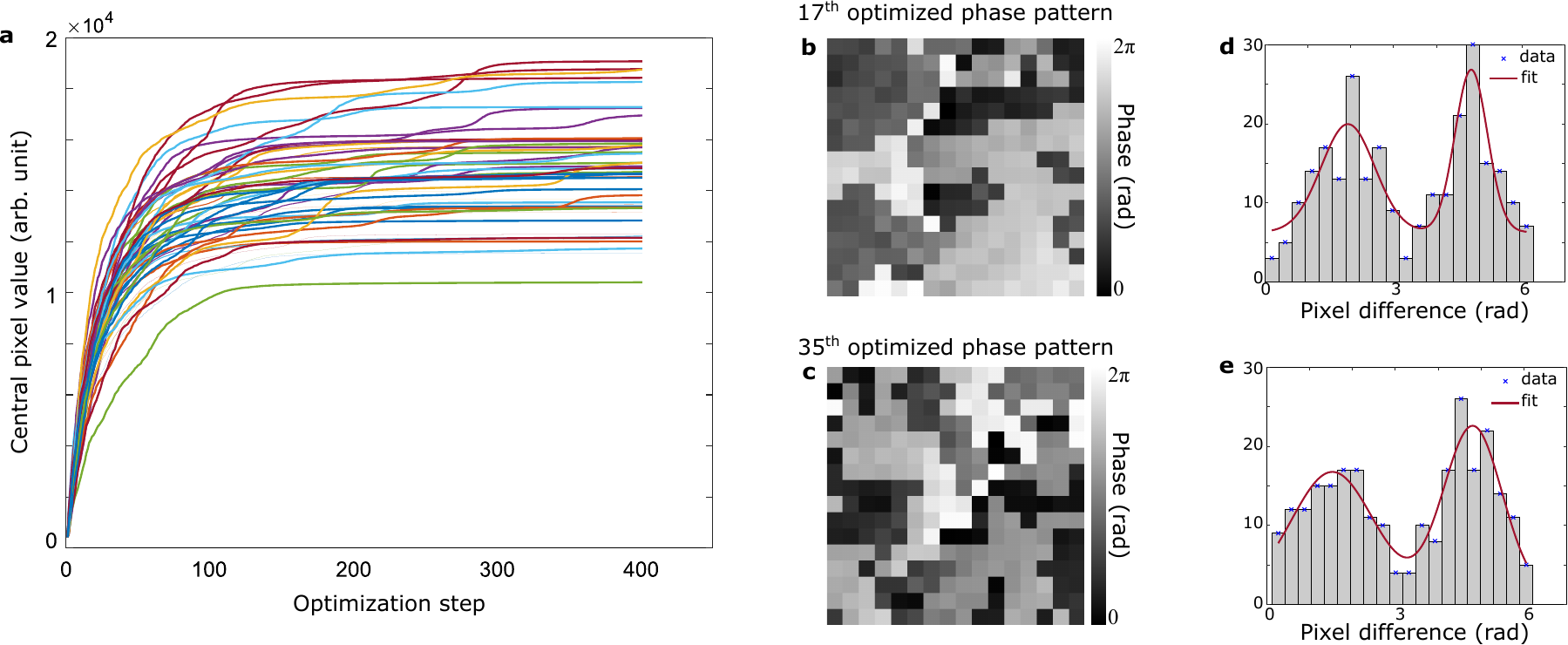}
    \caption{\textbf{{Different} existing non-trivial solutions $S'$}, {for} the random disorder {used in Figure 5a,e,i of the main document,} and a $16\cross 16$ SLM correcting pattern. \textbf{a}, Optimization curves for 50 different initial SLM phase patterns, {showing the variation of the central value in the correlation image in function of the number of steps. The optimization is performed computationally by simulating the experiment on a computer using experimentally measured transmission matrix and two-photon input state.} \textbf{b} and \textbf{c}, Optimized SLM pattern obtained for the $17^{th}$ and $35^{th}$ optimization, respectively. \textbf{d} and \textbf{e}, Histograms of the pixel-wise phase difference between the optimized phase pattern shown in \textbf{b} and \textbf{c} respectively, and the phase pattern derived from the scattering matrix implementing the identity transformation (see also Section~\ref{comparisonsolution}). The distribution are fitted by a double Gaussian distribution (red) centered on two values that are on average separated by $3.1\pm0.3$. }
    \label{fig:differentOpti}
\end{figure}

{In this section, we focus on the number of non-trivial solutions $S'$ that exist for a given configuration. As demonstrated by Equation (5) of the manuscript, which follows from the approximation $\sigma_{\vec{r}} \rightarrow 0$, there is a theoretical infinity of solutions for $S'$. However, since our experimental case does not exactly satisfy the assumption $\sigma_{\vec{r}} \rightarrow 0$, it is legitimate to ask whether the system always converges to the same solution $S'$ during the search for a solution by optimization.

To investigate this, we repeated the numerical optimization starting from different initial SLM phase masks. Figure~\ref{fig:differentOpti}a shows the optimization curves for 50 different initial SLM phase patterns. We observe that, while all optimizations reach a plateau, they do not converge to the same value. This behavior arises because multiple local maxima exist, as discussed in Ref.~\cite{courme2025nonclassicaloptimizationcomplexmedia}. These 50 phase patterns are all distinct pairwise, as the pairwise structural similarity is ranging from $-0.39$ to $0.48$, with an average of $0.001\pm 0.04$. They are also linearly independent, as they form a matrix of rank $50$.

Examining the optimized phase masks obtained in more detail, we see that they all differ from one another, as illustrated by the two examples shown in Figures~\ref{fig:differentOpti}b and c, which show the $17^{th}$ and $35^{th}$ optimized phase patterns, respectively. Although we cannot show all the patterns here, analyzing them reveals that they all are distinct from the SLM pattern calculated directly from the scattering matrix. This confirms that the system never converges to the trivial solution of reproducing the identity. Figure~\ref{fig:differentOpti}d and e show the distribution of pixel-wise phase differences between the optimized pattern and the SLM pattern implementing the identity transformation. In these two cases, as discussed in Section~\ref{comparisonsolution}, two distinct peaks separated by $\pi$ are clearly visible, confirming that the pattern is distinct from the trivial one. In fact, all the optimized patterns show the same behavior. Statistically, the histogram of the pixel-wise phase difference of among all the optimized patterns has two peaks, distant of $3.1\pm0.3$ from each other. This demonstrates that, while the optimization process cannot precisely direct the system towards the ideal solutions of $0$ and $\pi$ phases as predicted by Equation (5), it nonetheless converges towards solutions that closely resemble them. Therefore, we can conclude that multiple solutions for $S'$ also exist in the experimental case (i.e. $\sigma_{\vec{r}} > 0$).}

{\section{Details about the non-trivial solutions $S'$ and demonstration of Equation (5)} } 

In this section, we provide more details about the non-trivial solutions $S'$.

{\subsection{Verification of the solutions given by Equation (5)}}

Here, we outline the steps demonstrating that the solution $S'$ given by Equation (5) indeed enables the object to be reconstructed through the scattering medium using correlations.
We begin by establishing the general Equation (11) of the manuscript. The starting point is Equation (8), which describes the general propagation of a two-photon wavefunction from the input plane to the output plane:
\begin{equation}
\label{eq5SM}
\psi^{\text{out}}({\vec{r_i'}, \vec{r_s'}}) = \iint \psi^{\text{in}}({\vec{r_i}, \vec{r_s}}) s(\vec{r_i'}, \vec{r_i}) s(\vec{r_s'}, \vec{r_s}) d\vec{r_i} d\vec{r_s}.
\end{equation}
Then, we consider an input state of the form given by Equation (3) of the manuscript, whose derivation is detailed in Section~\ref{math3and10}. Inserting this input state into the above propagation equation and applying the change of variables $\vec{r}_+ = (\vec{r}_s + \vec{r}_i)/2$ and $\vec{r}_- = (\vec{r}_s - \vec{r}_i)/2$, we obtain:
\begin{eqnarray}
    \psi^{\text{out}}({\vec{r_i'}, \vec{r_s'}}) &=& \iint  {t} \left(\frac{\vec{r_i}+\vec{r_s}}{M} \right)  \exp{\frac{-|\vec{r_i}-\vec{r_s}|^2 \pi^2 \sigma_{\vec{r}}^2 }{4 \lambda_p^2 f_1^2}} s(\vec{r_i'}, \vec{r_i}) s(\vec{r_s'}, \vec{r_s}) d\vec{r_i} d\vec{r_s} \nonumber \\
    &=& \iint  {t} \left(\frac{2 \vec{r_+}}{M} \right)  \exp{\frac{- |\vec{r_-}|^2 \pi^2 \sigma_{\vec{r}}^2 }{ \lambda_p^2 f_1^2}} s(\vec{r_i'}, \vec{r_+}+\vec{r_-}) s(\vec{r_s'}, \vec{r_+}-\vec{r_-}) d\vec{r_+} d\vec{r_-} \nonumber \\
    \label{propEqgen} &=& \iint  {t} \left(\vec{r_+} \right) H_{s'}(\vec{r_i'},\vec{r_s'},\vec{r_+}) d\vec{r_+},
\end{eqnarray}
where $H_{s'}$ is defined by Equation (12) of the manuscript:
\begin{equation}
\label{eq9sm}
H_{s'}(\vec{r_i'}, \vec{r_s'}, \vec{r_+}) = \int s' \left(\vec{r_i'}, \frac{M \vec{r_+}}{2} +\vec{r_-} \right) s'\left (\vec{r_s'}, \frac{M \vec{r_+}}{2}-\vec{r_-} \right) \exp{\frac{- |\vec{r_-}|^2 \pi^2 \sigma_{\vec{r}}^2 }{ \lambda_p^2 f_1^2}} d\vec{r_-}.
\end{equation}
We then make the following assumption $\sigma_{\vec{r}} \approx 0$, which allows us to simplify the Gaussian term: $\exp{\frac{-|\vec{r_-}|^2 \pi^2 \sigma_{\vec{r}}^2 }{ \lambda_p^2 f_1^2}} \approx 1$ and retrieve Equation (15) of the manuscript. Next, we insert the solution given by Equation (5) of the manuscript into Equation (15):
\begin{eqnarray}
H_{s'}(\vec{r_i'}, \vec{r_s'}, \vec{r_+}) &=& \int s_g'(\vec{r_i'}, M \vec{r_+}/2 + \vec{r_-}) \, s_g'(\vec{r_s'}, M \vec{r_+}/2 - \vec{r_-}) \, d\vec{r_-} \nonumber \\
&=& \int \mathcal{F} \{ \operatorname{sign}(g) \}(\vec{r_i'}- M \vec{r_+}/2 - \vec{r_-}) \,\mathcal{F} \{ \operatorname{sign}(g) \}(\vec{r_s'} - M \vec{r_+}/2 + \vec{r_-}) \, d\vec{r_-} \nonumber \\
&=& \int \mathcal{F} \{ \operatorname{sign}(g) \}(\vec{R_i'} - \vec{r_-}) \,\mathcal{F} \{ \operatorname{sign}(g) \}(\vec{R_s'} + \vec{r_-}) \, d\vec{r_-} \text{ ,with } \vec{R_i'}=\vec{r_i'} - M \vec{r_+}/2 \text{ and } \vec{R_s'}=\vec{r_s'} - M \vec{r_+}/2 \nonumber \\
&=& \iiint e^{i (\vec{R_i'} - \vec{r_-}) \vec{k_i}}  \operatorname{sign}(g) (\vec{k_i}) \,e^{i (\vec{R_s'} + \vec{r_-}) \vec{k_s}} \operatorname{sign}(g)(\vec{k_s}) \, d\vec{r_-} d\vec{k_i} d\vec{k_s} \nonumber \\
&=& \iiint e^{i (\vec{k_i}-\vec{k_s}) \vec{r_-}} e^{i \vec{R_i'} \vec{k_i}}  e^{i \vec{R_s'} \vec{k_s}} \operatorname{sign}(g) (\vec{k_i}) \, \operatorname{sign}(g)(\vec{k_s}) \, d\vec{r_-} d\vec{k_i} d\vec{k_s} \nonumber \\
&=& \iint d\vec{k_i} d\vec{k_s}  e^{i \vec{R_i'} \vec{k_i}}  e^{i \vec{R_s'} \vec{k_s}}  \operatorname{sign}(g) (\vec{k_i}) \, \operatorname{sign}(g)(\vec{k_s}) \int e^{i (\vec{k_i}-\vec{k_s}) \vec{r_-}} \, d\vec{r_-} \nonumber \\
&=& \iint d\vec{k_i} d\vec{k_s}  e^{i \vec{R_i'} \vec{k_i}}  e^{i \vec{R_s'} \vec{k_s}}  \operatorname{sign}(g) (\vec{k_i}) \, \operatorname{sign}(g)(\vec{k_s}) \delta(\vec{k_i}-\vec{k_s})  \\
&=& \int d\vec{k}  e^{i (\vec{R_i'}+\vec{R_s'}) \vec{k}}  [\operatorname{sign}(g) (\vec{k})]^2   \nonumber \\
&=& \int d\vec{k}  e^{i (\vec{R_i'}+\vec{R_s'}) \vec{k}}   \nonumber \\
&=& \delta(\vec{R_i'}+\vec{R_s'})  \nonumber \\
&=& \delta \left( \vec{r_i'}+\vec{r_s'}-M \vec{r_+} \right).  
\end{eqnarray}
Finally, we insert the above expression of $H$ into Equation~\ref{propEqgen} to obtain:
$$
\psi^{\text{out}}(\vec{r_i'}, \vec{r_s'}) = t\!\left(\frac{\vec{r_i'} + \vec{r_s'}}{M}\right),
$$
which leads to the correlation image
$$
\Gamma^+(\vec{r_+}) =  \left| t\!\left(\frac{\vec{r_+}}{M}\right) \right|^2,
$$
thereby revealing the object.

{\subsection{Example of another encoding leading to a different class of non-trivial solutions $S'$.}}

In this subsection, we investigate an alternative form of image encoding in the correlations and show that there exists another class of non-trivial solutions that preserves the object image through a correlation measurement. In this case, we assume that the object image is encoded in the term $\vec{r_s} - \vec{r_i}$, as follows:
\begin{equation}
\label{eq3manuSM}
\psi^{\text{in}}({\vec{r_i},\vec{r_s}}) = {t} \left({\vec{r_i}-\vec{r_s}} \right) \exp{\frac{-|\vec{r_i}+\vec{r_s}|^2 \sigma_{\vec{k}}^2 }{4 }}.
\end{equation}
It should be noted that no experimental demonstration of this type of image encoding has been reported to date. Nevertheless, it could in principle be achieved by engineering the structure and geometry of the crystal rather than shaping the pump beam - a technique that has, for example, been employed in Ref.~\cite{yesharim_direct_2023}.
Assuming such encoding exists, one can show that, in the limit $\sigma_k \rightarrow 0$ — corresponding to a pump beam much larger than the crystal dimensions — functions of the form
\begin{equation}
\label{newsol}
   s_f'(\vec{r'}, \vec{r}) = \mathcal{F} \{ f \}(\vec{r'} - \vec{r}),
\end{equation}
where $f$ is an arbitrary complex function with the following property
\begin{equation}
    {\forall \vec{k}\in \mathbf{R}^3}, \, f(\vec{k}) f(-\vec{k})=1,
\end{equation}
constitute a class of non-trivial solutions for $S'$. To demonstrate this, we start again from Equation (8) of the manuscript, inserting the expression given in Equation~\ref{eq3manuSM}:
\begin{eqnarray}
    \psi^{\text{out}}({\vec{r_i'}, \vec{r_s'}}) &=& \iint  {t} \left(\vec{r_i}-\vec{r_s} \right)  \exp{\frac{-|\vec{r_i}+\vec{r_s}|^2  \sigma_{\vec{k}}^2 }{4}} s(\vec{r_i'}, \vec{r_i}) s(\vec{r_s'}, \vec{r_s}) d\vec{r_i} d\vec{r_s} \nonumber \\
    &=& \iint  {t} \left(2 \vec{r_-} \right)  \exp{{- |\vec{r_+}|^2  \sigma_{\vec{k}}^2}} s(\vec{r_i'}, \vec{r_+}+\vec{r_-}) s(\vec{r_s'}, \vec{r_+}-\vec{r_-}) d\vec{r_+} d\vec{r_-} \nonumber \\
    \label{propEqgen2} &=& \iint  {t} \left(\vec{r_-} \right) Q_{s'}(\vec{r_i'},\vec{r_s'},\vec{r_-}) d\vec{r_-},
\end{eqnarray}
where
\begin{equation}
\label{eq12sms}
Q_{s'}(\vec{r_i'}, \vec{r_s'}, \vec{r_+}) = \int s' \left(\vec{r_i'}, \frac{\vec{r_-}}{2} +\vec{r_+} \right) s'\left (\vec{r_s'}, \frac{-\vec{r_-}}{2}+\vec{r_+} \right) \exp{{- |\vec{r_+}|^2  \sigma_{\vec{k}}^2 }} d\vec{r_+}.
\end{equation}
We then make the following assumption $\sigma_{\vec{k}} = 0$, which allows us to simplify the Gaussian term: $\exp{{-|\vec{r_-}|^2 \sigma_{\vec{k}}^2 }} \approx 1$. Next, we insert the solution given by Equation~\ref{newsol} into the above Equation after simplifying the Gaussian term:
\begin{eqnarray}
Q_{s'}(\vec{r_i'}, \vec{r_s'}, \vec{r_+}) &=& \int s_f'(\vec{r_i'}, \vec{r_-}/2 + \vec{r_+}) \, s_f'(\vec{r_s'}, -\vec{r_-}/2 + \vec{r_+}) \, d\vec{r_+} \nonumber \\
&=& \int \mathcal{F} \{f \}(\vec{r_i'}- \vec{r_-}/2 - \vec{r_+}) \,\mathcal{F} \{ f \}(\vec{r_s'} + \vec{r_-}/2 - \vec{r_+}) \, d\vec{r_+} \nonumber \\
&=& \int \mathcal{F} \{ f \}(\vec{R_i'} + \vec{r_+}) \,\mathcal{F} \{ f \}(\vec{R_s'} + \vec{r_+}) \, d\vec{r_+} \text{ ,with } \vec{R_i'}=\vec{r_i'} - \vec{r_-}/2 \text{ and } \vec{R_s'}=\vec{r_s'} + \vec{r_-}/2 \nonumber \\
&=& \iiint e^{i (\vec{R_i'} + \vec{r_+}) \vec{k_i}}  f (\vec{k_i}) \,e^{i (\vec{R_s'} + \vec{r_+}) \vec{k_s}} f(\vec{k_s}) \, d\vec{r_+} d\vec{k_i} d\vec{k_s} \nonumber \\
&=& \iiint e^{i (\vec{k_i}+\vec{k_s}) \vec{r_+}} e^{i \vec{R_i'} \vec{k_i}}  e^{i \vec{R_s'} \vec{k_s}} f (\vec{k_i}) \, f(\vec{k_s}) \, d\vec{r_+} d\vec{k_i} d\vec{k_s} \nonumber \\
&=& \iint d\vec{k_i} d\vec{k_s}  e^{i \vec{R_i'} \vec{k_i}}  e^{i \vec{R_s'} \vec{k_s}}  f (\vec{k_i}) \, f(\vec{k_s}) \int e^{i (\vec{k_i}+\vec{k_s}) \vec{r_+}} \, d\vec{r_+} \nonumber \\
&=& \iint d\vec{k_i} d\vec{k_s}  e^{i \vec{R_i'} \vec{k_i}}  e^{i \vec{R_s'} \vec{k_s}} f(\vec{k_i}) \, f(\vec{k_s}) \delta(\vec{k_i}+\vec{k_s})  \\
&=& \int d\vec{k}  e^{i (\vec{R_i'}-\vec{R_s'}) \vec{k}}  f(\vec{k}) f(-\vec{k})    \nonumber \\
&=& \int d\vec{k}  e^{i (\vec{R_i'}+\vec{R_s'}) \vec{k}}   \nonumber \\
&=& \delta(\vec{R_i'}-\vec{R_s'})  \nonumber \\
&=& \delta \left( \vec{r_i'}-\vec{r_s'}-\vec{r_-} \right).  
\end{eqnarray}
Next, we insert the above expression for $Q$ into Equation~\ref{propEqgen2}, yielding:
$$
\psi^{\text{out}}(\vec{r_i'}, \vec{r_s'}) = t\left(\vec{r_i'} - \vec{r_s'}\right).
$$
To retrieve the encoded object, we then project the second-order correlation function $G^{(2)}$ onto the difference coordinate, defined as:
$$
\Gamma^-(\vec{r_-}) = \int G^{(2)}(\vec{r} + \vec{r_-}, \vec{r}) \, d\vec{r} = \left| t(\vec{r_-}) \right|^2.
$$
This projection reveals the object's intensity profile encoded in the correlations as a function of the coordinate difference $\vec{r_-}$.

\section{Deriving the classical solution for image transmission}  

The SLM phase pattern shown in Fig.~2b of the main manuscript, which corresponds to the classical solution and enables the implementation of a transformation $S'$ close to the identity, can be obtained either through classical optimization procedures~\cite{katz_looking_2012} or from the transmission matrix of the medium~\cite{gigan_matrice_2010}. In this work, we adopt the latter approach. The phase mask is directly obtained by extracting the phase of the vector resulting from the multiplication of the conjugate transpose of the experimentally measured transmission matrix $T$ (see Supplementary Section~\ref{matrixmeasure}) with a target output vector $E^{\mathrm{tar}}$, which contains zeros everywhere except for a single nonzero element at the position corresponding to the central pixel of the camera: 

\begin{equation}
    \Theta = \arg \left( T^\dagger E^{\mathrm{tar}} \right),
\end{equation}
where $\Theta$ is a vector containing the $32\times32$ phase values in $[0,2 \pi]$ programmed on the SLM. This approach is well established in the literature for transmitting images through scattering media~\cite{Bertolotti2022}. In practice, however, it is limited by the memory effect of the medium, which must be sufficiently large to provide an output field of view wide enough to reconstruct the image. As a result, this method works best for relatively thin scattering media. In our experiment, the parafilm layer is sufficiently thin to yield a field of view that fully encompasses the object.

\section{Details about the image transmission fidelity and the Figure 6a of the manuscript}

This section provides additional details on the image transmission fidelity values reported in Fig.~6a of the manuscript. These values are obtained by computing the structural similarity index (SSIM)~\cite{wang_image_2004} between an output image - either a classical intensity image or a quantum correlation image - after propagation through the system $S'$, consisting of the scattering medium followed by the spatial light modulator (SLM), and a reference image acquired in the absence of the scattering medium.

\begin{figure}[t!]
    \centering
\includegraphics[width=0.5 \linewidth]{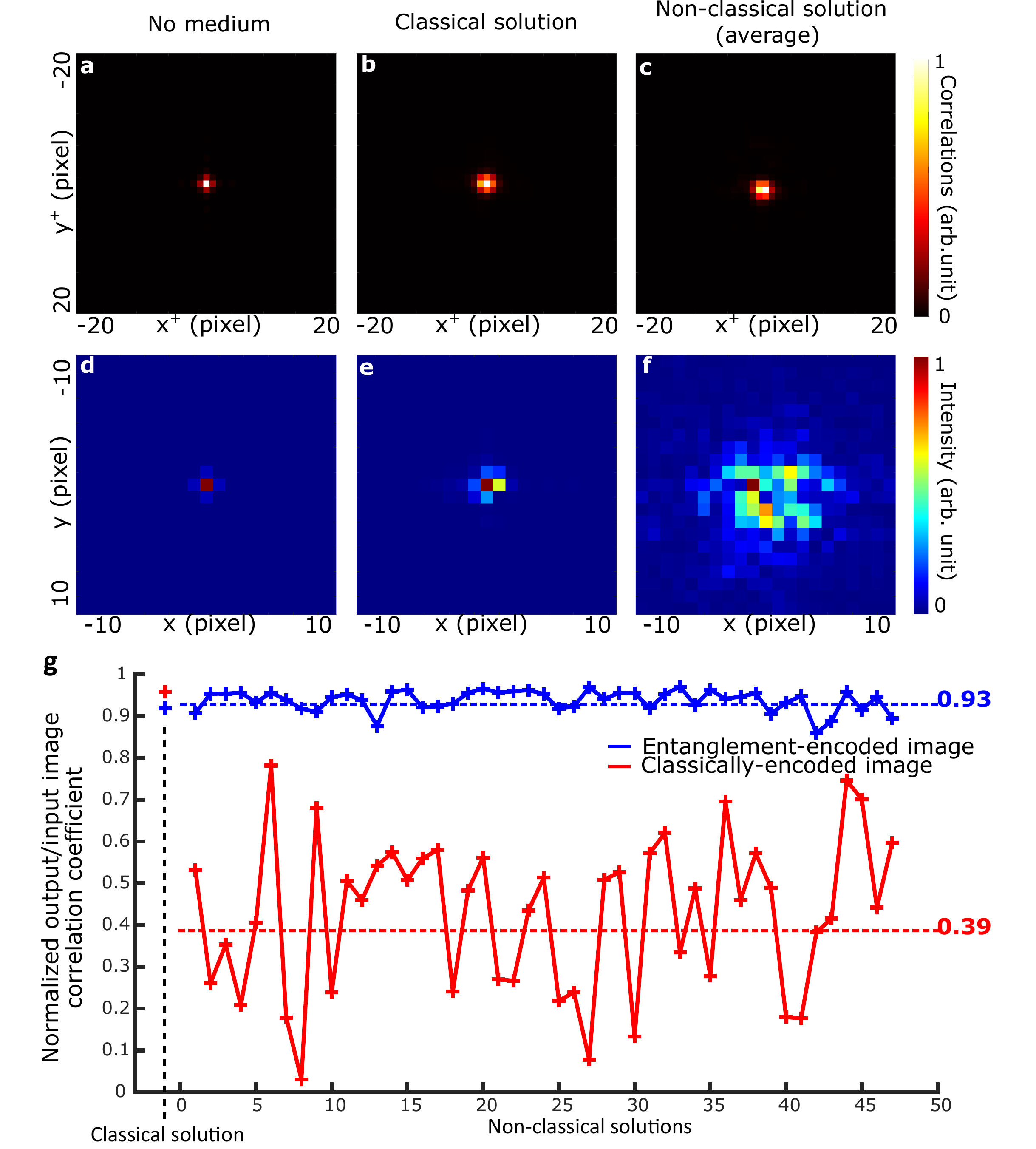}
    \caption{\textbf{Details of the image transmission fidelity.} \textbf{(a-f)} Images obtained from numerical simulations using an experimentally measured scattering matrix, from which the image transmission fidelity values shown in Fig.~6a are calculated. The input object is a point source. \textbf{(a)} Correlation image obtained without the scattering medium (reference). \textbf{(b)} Correlation image through the scattering medium using the classical solution. \textbf{(c)} Correlation image through the scattering medium using the non-classical solutions (average over 50 solutions). \textbf{(d)} Classical intensity image obtained without the scattering medium (reference). \textbf{(e)} Classical intensity image through the scattering medium using the classical solution. \textbf{(f)} Classical intensity image through the scattering medium using the non-classical solutions (average over 50 solutions). \textbf{(g)} Correlation values between the output image and the input image (reference without medium) for the classical solution and for 50 non-classical solutions. In this case, the correlation values are used to estimate the image transmission fidelity. The scattering matrix is generated purely numerically, and the input object is the digit ``8''. The blue curve corresponds to results obtained with an entangled state, while the red curve corresponds to a classical state. {The average transmission fidelity
over the 50 non-classical patterns is $0.93 \pm 0.03$ for the entangled state (blue line) and $0.39 \pm 0.24$ for the classical encoding (red line).}}
    \label{fig:comparison}
\end{figure}

The output images are generated through numerical simulations using a point source as the object and an experimentally measured transmission matrix of the scattering medium. The same transmission matrix is used to produce the results shown in Fig.~3 of the manuscript. Representative examples of these images are shown in Figs.~\ref{fig:comparison}a-f.

The transmission fidelity is defined by a normalized SSIM,
\[
(\mathrm{SSIM}-\mathrm{SSIM}_0)/(1-\mathrm{SSIM}_0),
\]
where $\mathrm{SSIM}_0$ denotes the mean SSIM between the reference image and images obtained after propagation through the scattering medium when random phase patterns are applied to the SLM.

In addition to these results, we performed supplementary numerical simulations, the results of which are shown in Fig.~\ref{fig:comparison}g. In this case, the scattering matrix $T$ of the medium is fully simulated (see Section~\ref{generaldetails}), and the input object is the digit ``8''. Furthermore, we evaluate the transmission fidelity using an alternative metric based on the correlation between the input and output images. In practice, this metric is computed in \textsc{Matlab} as
\[
\mathrm{C} = \max\!\left(\max\!\left(\mathrm{xcorr2}(I,O)\right)\right),
\]
where $I$ and $O$ denote the input and output images, respectively. As for the SSIM, the correlation metric is normalized according to
\[
(\mathrm{C}-\mathrm{C}_0)/(1-\mathrm{C}_0),
\]
where $\mathrm{C}_0$ is the mean correlation value obtained for images transmitted through the scattering medium using random phase patterns on the SLM.

The resulting values - in particular the ratio between the transmission fidelity in the quantum case (blue curve) and in the classical case (red curve) - are very similar to those reported in Fig.~6a of the manuscript. This agreement further confirms the effective filtering enabled by our approach.

{\section{Existence of multiple non-classical Solutions for arbitrary entangled two-photon input states $\Psi^{\text{in}}$ and complex media} 

We consider a pure two-photon state characterized by a symmetric matrix
$\Psi^{\textbf{in}}$, which propagates through a scattering medium described
by a unitary matrix $S$. The two-photon output state is then given by
\begin{equation}
\label{eqSM1}
    \Psi^{\textbf{out}} = S \Psi^{\textbf{in}} S^{t}.
\end{equation}

We now assume that an ideal wavefront-shaping device is inserted between the source and the
scattering medium. This device is assumed to be capable of
implementing an arbitrary unitary transformation $Q$. The propagation equation
then becomes
\begin{equation}
\label{eqSM1bis}
    \Psi^{\textbf{out}} = S Q \Psi^{\textbf{in}} (S Q)^{t}.
\end{equation}

Although such a device does not yet exist for an arbitrarily large number of
modes, it is currently under development and has already been demonstrated
experimentally for a small number of modes, notably for imaging through
multimode fibers~\cite{kupianskyi_all-optically_2024}. In this section, we show that for any input
state $\Psi^{\textbf{in}}$, there always exists at least one matrix $Q$,
distinct from the trivial solution $S^{-1}$, that leaves the input state
invariant i.e.
\[
\Psi^{\textbf{out}} = \Psi^{\textbf{in}},
\]
and therefore also preserves the associated correlations,
$|\Psi^{\textbf{out}}|^2 = |\Psi^{\textbf{in}}|^2$.

\subsection{Schmidt decomposition}

The first result we use is the Schmidt decomposition of the (pure) input state.
It allows us to write
\begin{equation}
    \Psi^{\textbf{in}} = T D_{\textbf{in}} T^{t},
\end{equation}
where $T$ is a unitary matrix and $D_{\textbf{in}}$ is a diagonal matrix with
$p$ non-zero, real, and positive diagonal elements
$\sqrt{\lambda_i}$, with $i \in [1,p]$. The $\lambda_i$ are the Schmidt
coefficients of the state.
Here, the fact that $\Psi^{\textbf{in}}$ is symmetric (since the two photons
share the same frequency and polarization) allows the decomposition to be
written with the same matrix $T$ on the left and its transpose on the right.
This is also known as the \emph{Takagi decomposition}. \cite{defienne_general_2018}

\subsection{Invariance under multiplication by $\pm 1$}

The second result is a simple property of matrix multiplication. Let $D$ be an
arbitrary diagonal matrix, and let $D_{1/-1}$ be a diagonal matrix whose entries
are randomly chosen from $\{+1,-1\}$. Then the following identity holds:
\begin{equation}
    D_{1/-1} \, D \, (D_{1/-1})^{t} = D.
\end{equation}

\subsection{Proof of the existence of $Q$}

We now return to the propagation equation~\eqref{eqSM1bis} and consider the
specific choice
\begin{equation}
\label{eqQ}
   Q = S^{-1} T D_{1/-1} T^{-1} .
\end{equation}
For this choice of $Q$, we obtain
\begin{eqnarray}
\label{eqSM5}
    \Psi^{\text{out}}
    &=& S Q \Psi^{\text{in}} (S Q )^t \nonumber \\
    &=& T D_{1/-1} T^{-1} \Psi^{\text{in}}
        \left( T D_{1/-1} T^{-1} \right)^t \nonumber \\
    &=& T D_{1/-1} T^{-1} \Psi^{\text{in}} (T^{t})^{-1}
        D_{1/-1} T^{t} \nonumber \\
    &=& T D_{1/-1} D_{\textbf{in}} D_{1/-1} T^{t} \nonumber \\
    &=& T D_{\textbf{in}} T^{t} \nonumber \\
    &=& \Psi^{\text{in}}.
\end{eqnarray}
This choice of $Q$ therefore leaves the input state invariant.
It remains to verify that $Q$ is not equal to the trivial solution $S^{-1}$.
From Eq.~\eqref{eqQ}, we see that multiple choices of $Q$ are possible,
depending on the positions of the $\pm 1$ entries in $D_{1/-1}$ and on the
dimension of the system. The cases where all diagonal elements are $+1$ or all
$-1$ correspond to the trivial solution $Q = S^{-1}$ (up to a global phase).
Importantly, as soon as the system dimension is greater than or equal to $2$,
it is always possible to choose $D_{1/-1}$ such that $Q \neq S^{-1}$.

\subsection{System dimension and entanglement}

The existence of at least one non-trivial solution is therefore guaranteed as
soon as the system dimension is greater than or equal to $2$. Notably, the
dimension of the system is precisely given by the number of non-zero diagonal
elements of $D_{\textbf{in}}$, i.e., by definition, the number of non-zero
Schmidt coefficients.
Consequently, the only case for which no non-trivial solution exists is when
there is a single non-zero Schmidt coefficient, corresponding to a separable
state. As soon as the input state is entangled, at least one non-classical solution
necessarily exists, thereby concluding the proof.

\subsection{Number of solutions}

For a system of dimension $p$ (i.e. with $p$ non-zero Schmidt coefficients),
there exist $2^{p}$ distinct arrangements of $\pm 1$ on the diagonal of
$D_{1/-1}$. After removing redundant cases corresponding to a global $\pi$
phase and excluding the trivial solution $Q = S^{-1}$, we obtain a total of
$2^{p-1}-1$ distinct non-classical solutions.

It is important to emphasize, however, that this number does not represent the
complete set of non-classical solutions that preserve correlations. Indeed, in
order to prove existence, we have imposed a stronger condition than mere
correlation preservation, namely the full invariance of the two-photon state.
This restriction likely excludes additional solutions that preserve
correlations without leaving the state strictly invariant.}

%